\documentclass[runninghead,10pt,times]{llncs}
\usepackage{graphicx,subfigure,times,amsmath,amssymb} 
\usepackage{url,footnote,threeparttable,algorithm}
\usepackage[square, comma, sort&compress]{natbib}
\usepackage{epsfig}
\usepackage{multirow}
\usepackage{caption3}
\usepackage{verbatim}
\usepackage{algorithmic,algorithm}

\newcommand{\keywords}[1]{\par\addvspace\baselineskip
\noindent\keywordname\enspace\ignorespaces#1}

\makeatletter
\renewcommand\bibsection%
{
  \section*{\refname
    \@mkboth{\MakeUppercase{\refname}}{\MakeUppercase{\refname}}}
} \makeatother

\begin{document}
%
\title{Securing Anonymous Communication Channels under the Selective DoS Attack}
\author{
Anupam Das \and Nikita Borisov}
\institute{University of Illinois at Urbana Champaign, USA\\
\email{\{das17,nikita\}@illinois.edu}\\
}

\maketitle

\begin{abstract}
Anonymous communication systems are subject to selective denial-of-service (DoS) attacks. Selective DoS attacks lower
anonymity as they force paths to be rebuilt multiple times to ensure delivery which increases the opportunity for more
attack. In this paper we present a detection algorithm that filters out compromised communication channels for one of
the anonymity networks, Tor. Our detection algorithm uses two levels of probing to filter out potentially compromised
tunnels. We perform probabilistic analysis and extensive simulation to show the robustness of our detection algorithm.
We also analyze the overhead of our detection algorithm and show that we can achieve satisfactory security guarantee
for reasonable communication overhead ($\sim5\%$ of the total available Tor bandwidth). Real world experiments reveal
that our detection algorithm provides good defense against selective DoS attack. \keywords{Anonymity, Tor network,
denial of service (DoS) attack.}
\end{abstract}
\allowdisplaybreaks

\section{Introduction}
Anonymous communication was first introduced in 1981 with Chaum's seminal paper on ``Untraceable electronic mail,
return addresses, and digital pseudonyms'' \cite{Chaum}. Since then, many researchers have concentrated on building,
analyzing and attacking anonymous communication systems like Tor~\cite{Dingledine2004}, I2P~\cite{I2P} and
Freenet~\cite{Freenet}. In this paper we concentrate on Tor~\cite{Dingledine2004}, one of the most widely used
low-latency anonymity networks, which conceals users' identities and activities from surveillance and traffic analysis.
Tor provides confidentiality and privacy to users of various types ranging from ordinary individuals to business
personnel, journalists, government employees and even military personnel. Currently, Tor has over 3000 relays all
around the world and it provides anonymity to hundreds of thousands of people every day
\cite{Hahn2010,Loesing2009,torbandwidth}.

Users' identity, however, can become exposed when multiple relays are compromised. By default, Tor uses three relays
and an attacker who can gain control of the entry and exit relays is capable of compromising user identity using timing
analysis \cite{Neil2004,Shmatikov2006}. Moreover, malicious nodes can perform a selective denial-of-service (DoS)
attack \cite{Borisov2007,Bauer2007} where compromised relays drop packets until a path starting and ending with a
compromised node is built. This increases the probability of such a path being built and as a result lowers anonymity.
Recently, the Dutch ministry of Justice and Security proposed passing a law which would enable the law enforcement
office to launch any form of attack (i.e., selective DoS could one form of attack) on any system in order to gather
evidence \cite{dutch}. So some form of mechanism is needed to ensure secure path construction in the presence of
compromised/controlled relays.

Danner et al.~\cite{Danner2009} showed that it is possible to identify relays mounting selective DoS using exhaustive
probing. The intent is to periodically carry out these probes and blacklist the misbehaving relays; however, the total
number of probes required is prohibitive---3 times the size of the network (several thousand routers in Tor) , and many
more (typically retrying each probe 10 times) to account for non-malicious failures. So, their approach seems practical
for a centralized design, but we wanted to create a local mechanism to defend against selective DoS. By using
probabilistic inference, we can make do with orders of magnitude fewer probes and thus our approach can be practical to
be run at each individual client.

Like Danner et al., we make use of probing, but rather than deterministically identifying all relays that are
performing selective DoS, we probabilistically check the safety of particular circuits. In particular, a user builds a
number of circuits for future use and then evaluates their safety by checking whether nodes involved in a circuit will
perform a selective DoS attack. Our tests are based on the assumption that only a minority of Tor nodes are
compromised; this is generally assumed to be necessary to receive reasonable anonymity protection from Tor, regardless
of selective DoS attacks.

\textbf{Contributions.} We make the following contributions: $(a)$ We present a detection mechanism to filter
potentially compromised communication channels. $(b)$ We provide a probabilistic model of our detection mechanism. Our
algorithm ensures that an attacker who performs selective DoS is unsuccessful at compromising tunnels with high
probability. $(c)$ We investigate adaptive attackers who change their strategy specifically in response to our
detection scheme. We find that, depending on the choice of parameters, the dominant strategy for such attackers is to
not perform selective DoS. $(d)$ We perform extensive simulation and some real world experiments to show the
effectiveness of our detection mechanism.

\textbf{Roadmap.} Section {\ref{background}} gives an overview of Tor network along with the threat model. In Section
{\ref{model}} we formally introduces our algorithm. We provide security analysis of our detection algorithm in Section
{\ref{analysis}}. To show the effectiveness of our algorithm, we present simulation and experimental results in Section
{\ref{evaluation}}. We describe some of the related works in Section {\ref{related_work}}. Finally we conclude in
Section {\ref{conclusion}}.

\section{Background}{\label{background}}

\subsection{Tor Network}
Tor~\cite{Dingledine2004} is an anonymous communication network that allows users to make TCP connections to Internet
sites without revealing their identity to the destination or third-party observers.  We will briefly explain the main
components of the operation of Tor as they are relevant to this work. To initiate an anonymous TCP connection, a Tor
user constructs a \emph{circuit} (also known as a tunnel or path) comprised of several Tor \emph{relays} (also known as
routers). The relays form a forwarding chain that sends traffic from the user to the destination, and vice versa.
Circuits typically involve three relays: the \emph{entry}, \emph{middle}, and \emph{exit}. The traffic contents are
protected by a layered encryption scheme, called onion routing~\cite{Reed1998}, where each relay peels off a layer
while forwarding. As a result, any individual router cannot reconstruct the whole circuit path and link the source to
the destination.
The relays in a circuit are chosen using specific constraints~\cite{TorPathSpec}.  Each user selects a small, fixed
number of entry relays that are used for all circuits. 
These relays are called \emph{guard relays}~\cite{overlier-syverson:oakland06,Wright2003}; their use is designed to
defend from the predecessor attack~\cite{Wright2002}. To choose the exit relay, the user picks from among those relays
that have an exit policy
compatible with the desired destination.  After
these constraints, the relays for each position are chosen randomly, weighted by their bandwidth.

Tor aims to provide low-latency traffic forwarding for its users. As a result, as traffic is forwarded along the path
of a circuit, timing patterns remain discernible, and an attacker who observes two different relays can use timing
analysis to determine whether they are in fact forwarding the same
circuit~\cite{Neil2004,Shmatikov2006,Syverson2001,Zhu2004}. As a result, to link a Tor user to a destination, it
suffices to observe the entry and the exit relays of a circuit.  Standard security analysis of
Tor~\cite{Syverson2001,Dingledine2004} shows that if $t$ is the fraction of relays that are observed, an adversary will
be able to violate anonymity on $t^2$ of all of the circuits. Note that, due to  bandwidth-weighted path selection in
Tor, $t$ is best thought of as the fraction of total Tor \emph{bandwidth} that belongs to relays under
observation\footnote{To be more precise, the correct fraction would be $t_g \cdot t_e$, where $t_g$ and $t_e$ are the
fractions of the guard and exit bandwidth under observation, respectively.  For simplicity of presentation, we will
assume $t_g = t_e = t_m = t$ in the rest of the paper.}.  The security of Tor, therefore, relies on the assumption that
a typical adversary will not be able to observe a significant fraction of Tor relays.  For most adversaries, the
easiest way to observe relay traffic is to run their own relays (i.g., using cloud infrastructures like Amazon's
EC2,Rackspace etc.) or compromise existing ones.

\subsection{Selective Denial of Service in Tor}\label{SDoS}
An adversary who controls a Tor relay can perform a number of active attacks to increase the odds of
compromise~\cite{Borisov2007,Bauer2007}. One approach, which is the focus of this work, is \emph{selective denial of
service}~\cite{Borisov2007}. A compromised relay that participates in a circuit can easily check whether both the entry
and exit relays are under observation. If this is not the case, the relay can ``break'' the circuit by refusing to
forward traffic. This will cause a user to reformulate a new circuit for the connection, giving the adversary another
chance to compromise the circuit.  A simple analysis shows that this increases the overall fraction of compromised
circuits to $\frac{t^2}{t^2 + (1-t)^3} > t^2$, because only circuits with compromised entry and exit relays ($t^2$) or
circuits with no compromised relays ($(1-t)^3$) will be functional, and out of those $t^2$ will be compromised. E.g.,
if $t=0.2$, selective DoS increases the fraction of compromised circuits from 4\% to 7.25\%. The use of guard nodes
changes the analysis somewhat; compromised guards can amplify the effect of selective DoS. Bauer et al.
\cite{Bauer10onthe} showed that deploying a moderate number of inexpensive\footnote{Middle-only nodes do not have to
fulfill stronger commitments (e.g., minimum bandwidth, minimum uptime, legal issues related to exit policies) that
guard and exit nodes have to fulfill.} middle-only relays can boost the effect of selective DoS attack.

\subsection{Threat Model}
In our threat model we assume that a small fraction (typically 20\%) of the Tor relays are compromised and compromised
relays carry out selective DoS attack. Compromised relays may choose to perform probabilistic dropping, where a
compromised relay terminates a certain fraction of all circuits that it cannot compromise. We also discuss other clever
strategies that can be adopted against our detection algorithm in Section \ref{strategy}. Finally, we assume that
probes are indistinguishable from real user traffic. We describe different ways to achieve this in Section
\ref{masking-probes}.

\section{Our Detection Algorithm}{\label{model}}

\subsection{Overview}
Our algorithm is built on the fundamental assumption of the Tor security model that a relatively small fraction of all
relays are compromised. The algorithm works in two phases and runs periodically. In the first phase, we construct
random circuits following the Tor path construction algorithm and then test their functionality. Under a selective DoS
attack, we expect only two types of circuits to work: \emph{honest} circuits that contain no malicious relays, and
\emph{compromised} circuits with compromised entry and exit relays. In the second phase, we cross-check the circuits
against each other by changing their exit relays (and optionally middle relays too). The prevalence of honest circuits
means that compromised circuits will encounter more failures in the second phase than honest ones and therefore can be
identified. Table \ref{tab1} summarizes the different parameters used for our detection algorithm.

\begin{table}
\centering \caption{Parameters Used}\resizebox{7cm}{!}{
\begin{tabular}{|c|c|l|}
\hline
Setting&Parameter&Description\\
\hline
&t&Fraction of relays compromised\\
\cline{2-3}
Environmental&g&Fraction of compromised guards per user\\
\cline{2-3}
&f&Random network failure\\
\cline{2-3}
&d&Random drop rate by compromised nodes\\
\hline
&N&\# of working Tor circuits created in 1st phase\\
\cline{2-3}
Tunable&K&\# of probes used per circuit in 2nd phase\\
\cline{2-3}
&\emph{Th}&Threshold for classifying circuit\\
\hline
\end{tabular}}
\label{tab1}
\end{table}

\subsection{First Phase}
Tor circuits live for 10 minutes, meaning that we need 6 non-compromised circuits every hour. So in the first phase of
our detection algorithm we iteratively generate a random Tor circuit (following Tor path
specification~\cite{TorPathSpec}) and test its functionality by retrieving a random web file through the circuit. If it
fails we discard the circuit and try a new circuit. We stop when we have $N$ (in Section \ref{tuning} we will show how
to calculate the value of $N$) working circuits. If an adversary is carrying out selective DoS attack then after the
first phase we should have a set of circuits of form either \emph{CXC} or \emph{HHH}, where \emph{C} denotes a
compromised relay, \emph{H} denotes an honest one, and \emph{X} is a relay of any type. Note that some circuits of the
above forms may still fail for ``natural'' reasons, such as an overloaded relay. We discuss the impact of such failures
in Section \ref{failures}.

\subsection{Second Phase}
\begin{figure}[!h]
\centering
\includegraphics[width=0.6\columnwidth]{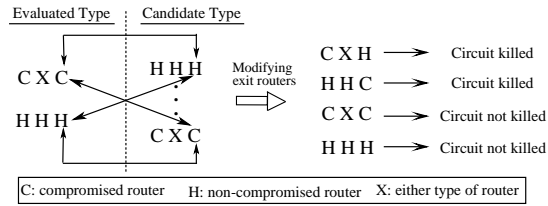}
\caption{Modifying the of exit relay of the \textit{evaluatee} with
that of the \textit{candidate} circuit to generate a probing circuit
in the second phase.} \label{fig2}
\end{figure}

In the second phase, we examine each of the circuits passing the first phase (we will call these circuits as
\emph{potential} circuits) by cross-checking them with each other. We evaluate each \emph{potential} circuit as
follows:\nolinebreak
\begin{itemize}
\item[$\bullet$] We randomly pick $K (1\leq K <N)$ other circuits (we will call them as \emph{candidate} circuits) out of
the list of potential circuits.
\item[$\bullet$] For each of the $K$ \emph{candidate}\footnote{Tor circuits have to
follow the constraint that no two relays in the same circuit can be within the same /16 network, or part of the same
operator-specified family. Therefore, when choosing the $K$ candidate circuits, we pick only those whose exit nodes
that don't violate this constraint.} circuits, we change the exit relay of the \emph{potential} circuit being evaluated
with the exit relay of the candidate circuit (a schematic description is shown in Figure~\ref{fig2}). We then test the
functionality/reliability of the new circuit by performing a test retrieval through it.  If, out of these $K$ probes,
$\mathit{Th}$ or more succeed, we consider the evaluated circuit to be honest; otherwise, we consider it to be
compromised. We also propose selecting random middle nodes to make probes indistinguishable from actual circuits in
Section \ref{masking-probes}.
\end{itemize}
Note that under selective DoS, if we change the exit relay of a compromised circuit with that of an honest circuit, we will get a circuit where the
entry is compromised and the exit is honest and hence the file retrieval should fail.  On the other hand, if the evaluated and candidate circuits are
both honest, or both compromised, the probe will succeed. We expect more success for an honest circuit, since most of the potential circuits are
honest; we use $\mathit{Th}$ as a threshold for distinguishing between the two circuit types. At the end of second phase, we will have some number of
potentially honest circuits. This collection of circuits is then used for making real anonymous connections. Once the pool of circuits is exhausted
(typically after one hour if used continuously), the algorithm is run again to identify new honest circuits (ideally, the algorithm would be started
ahead of time to avoid a delay in circuit availability). The pseudo-code of our detection algorithm is given in Algorithm~\ref{alg1}.

\begin{algorithm}
\caption{Generating secured Tor circuits} \label{alg1}
\begin{algorithmic}
\STATE  {\bf Input:} {List of stable, valid and running Tor routers. Parameters $N$, $K$ and $\mathit{Th}$}

\STATE {\bf Output:} {List of usable Tor circuits}

\STATE {\underline{1st Phase:}}

\STATE {$i \leftarrow 1$}

\WHILE {$i\leq N$}
    \STATE {Create a random Tor circuit according
to Tor path specification}

    \STATE {Probe the Tor circuits to check its reliability}

    \IF {probing successful}

        \STATE {$i \leftarrow i+1$}

    \ENDIF

\ENDWHILE

\STATE {Consider the set of circuits that pass this phase as $P$}

\STATE {\underline{2nd Phase:}}

\STATE {$S \leftarrow \emptyset$}

\FOR {$each$ $x \in P$}

    \STATE {Choose $K$ other circuits from $P$ such that they do not
violate Tor path constraints}

    \STATE {$count \leftarrow 0$}

    \FOR {$each$ $y \in K$}

        \STATE {Modify the exit relay of $x$ with that of $y$}

        \STATE {Probe the modified circuit}

        \IF {probing successful}

            \STATE {$count \leftarrow count+1$}

        \ENDIF

    \ENDFOR

    \IF {$count \geq Th$}

        \STATE {Classify $x$ as honest circuit}

        \STATE {$S \leftarrow S \cup x$}

    \ELSE

        \STATE {Classify $x$ as compromised circuit}

    \ENDIF

\ENDFOR

\STATE {\bf {return} $S$}
\end{algorithmic}
\end{algorithm}

\section{Security Analysis}{\label{analysis}}
In this section we will analyze the various security aspects of our detection algorithm through a probabilistic model.

\subsection{Considering Compromised Guard Nodes}{\label{guard_node}}
The selective DoS attack is particularly troubling due to the predecessor attack \cite{Wright2002}. To resist the
predecessor attack, Tor introduced a defensive mechanism where each user picks a small fixed set of guard nodes
(usually 3) that are always used as the first node in a circuit \cite{Wright2003}. As a result, users who pick honest
guard nodes will be immune to the predecessor attack. Guard nodes similarly help defend against selective DoS where
users with honest guards will never have their anonymity compromised. However, in another way, if certain fraction of
the guard nodes are compromised it can actually make the selective DoS attack more powerful. Assume a user has $G$
(typically $G=3$) guard nodes where $g$ fraction of them are compromised. So $g\in\{0,1/3,2/3,1\}$ for different number
of compromised guard nodes.

\subsection{False Error Rates}
We will first analyze the false-negative and false-positive rates of our algorithm under selective DoS strategy. For
now we assume that the adversary terminates any circuit which it cannot compromise; we consider other clever strategies
in Section~\ref{strategy}.

We consider an adversary who compromises a bandwidth-weighted fraction $t$ of all Tor relays. First, we will calculate
false-negative (FN) rate--the fraction of compromised circuits that pass our detection algorithm. The probability of
$FN$ depends upon the number of compromised ($\mathit{CXC}$) and honest ($\mathit{HHH}$) circuits randomly created in
the first phase. Let, $c$ and $h$ represent the number of circuits of the form $\mathit{CXC}$ and $\mathit{HHH}$
respectively. Since the adversary is doing selective dropping the only circuits that pass the first phase of our
algorithm are either compromised or honest, so $c+h=N$. Then probability of false negative error can be calculated by
the following function: \nolinebreak
\begin{equation}
\Pr(\mathit{FN})=\sum_{c=0}^{N} \Pr(C=c)\Pr(\mathit{FN}|C=c,H=N-c)\label{eqn4}
\end{equation}
\noindent where $C$ and $H$ are the numbers of compromised and honest circuits after phase 1, respectively. Note that,
likewise, for false-positive errors (fraction of honest circuits falsely discarded): \nolinebreak
\begin{equation}
\Pr(\mathit{FP})=\sum_{c=0}^{N} \Pr(C=c)\Pr(\mathit{FP}|C=c,H=N-c)
\end{equation}
To calculate the individual probabilities in~\eqref{eqn4}, we can approximate $C$ by the Binomial distribution
$B(N,\frac{gt}{gt+(1-g)(1-t)^2})$ and since $c+h=N$, $\Pr(H=N-c|C=c)=1$. This calculation assumes that relays are
sampled \emph{with} replacement and that family and /16 subnet constraints are ignored, but given the large number of
Tor relays, this results in minimal approximation error.

A false-negative error occurs when a compromised circuit is paired with at least $\mathit{Th}$ other compromised
candidate circuits. Since these circuits are sampled without replacement, we can calculate $\Pr(\mathit{FN}|C=c,H=N-c)$
using a hypergeometric distribution: \nolinebreak
\begin{equation}
\Pr(\mathit{FN}|C=c,H=N-c)=\sum\limits_{i=Th'}^{K'}\frac{\binom{c-1}{i}\binom{N-c}{K'-i}}{\binom{N-1}{K'}}\label{eq:fn}
\end{equation}
\noindent where $K' = \min(K,N-1)$ and $Th' = \min(Th,K')$. Similarly, a false-positive error occurs when an honest
circuit is paired with fewer than $\mathit{Th}$ honest candidate circuits: \nolinebreak
\begin{equation}
\Pr(\mathit{FP}|C=c,H=N-c)=\sum\limits_{i=0}^{Th'-1}\frac{\binom{N-c-1}{i}\binom{c}{K'-i}}{\binom{N-1}{K'}}\label{eq:fp}
\end{equation}

We also analyze the impact of transient network failures on both FN and FP rates on the next section.

\subsection{Dealing with Transient Network Failures}{\label{failures}}
Regardless of whether selective DoS is being performed or not, circuit failures can and will happen in Tor. This may be caused by connectivity errors
in the network, or, more likely, congestion at or crash of the Tor relay. In any network infrastructure there is always some network failure. Since,
network failure can directly influence the success rate of our probing, it can affect both $FN$ and $FP$ rates. So equation (\ref{eqn4}) needs to be
updated. After phase 1, we have to consider that a fraction of both $CXC$ and $HHH$ circuits fail. Lets assume that out of the $C$, \emph{CXC} and
$H$, \emph{HHH} circuits, $C'$ and $H'$ respectively survive random network failure. Then $\Pr(FN)$ can be computed using equation
(\ref{eqn:failures-included}). Terms $\Pr(C'=c'|C=c)$ and $\Pr(H=h'|H=N-c)$ can be approximated by binomial distributions $B(c,c',1-f)$ and
$B(N-c,h',1-f)$ respectively (where $f$ denotes random failure rate). So after considering random network failure we have $c'$ compromised and $h'$
honest circuits during the second phase. We also have to consider failures in the second phase, where certain fraction of the modified circuits will
fail to retrieve a file through the network. This will cause equation (\ref{eq:fn}) and (\ref{eq:fp}) to be modified to equations
(\ref{eqn:hyper-failure1}) and (\ref{eqn:hyper-failure2}). In equations (\ref{eqn:hyper-failure1}) and (\ref{eqn:hyper-failure2}), $K' =
\min(K,c'+h'-1)$ and $Th' = \min(Th,K')$.\nolinebreak
\begin{align}
\Pr(FN)&=\sum\limits_{c=0}^{N}\Pr(C=c)\left[\sum\limits_{c'=0}^{c}\Pr(C'=c'|C=c)\Pr(FN|C'=c',C=c)\right]\nonumber\\
&=\sum\limits_{c=0}^{N}\Pr(C=c)\left[\sum\limits_{c'=0}^{c}\Pr(C'=c'|C=c)\left[\vphantom{\sum\limits_{h=0}^{N-c}}\sum\limits_{h'=0}^{N-c}\Pr(H'=h'|H=N-c)\right.\right.\nonumber\\
&\hspace{0.75in}\left.\left.\vphantom{\sum\limits_{h=0}^{N-c}}\cdot\Pr(FN|C=c,C'=c',H=N-c,H'=h')\right]\right] \label{eqn:failures-included}
\end{align}
\nolinebreak
\begin{align}
&\Pr(\mathit{FN}|C'=c',H'=h')=\sum\limits_{i=Th'}^{K'}\frac{\binom{c'-1}{i}\binom{h'}{K'-i}}{\binom{c'+h'-1}{K'}}\sum\limits_{j=Th'}^{i}B(i,j,1-f)\label{eqn:hyper-failure1}\\
&\Pr(\mathit{FP}|C'=c',H'=h')=1-\sum\limits_{i=Th'}^{K'}\frac{\binom{h'-1}{i}\binom{c'}{K'-i}}{\binom{c'+h'-1}{K'}}\sum\limits_{j=Th'}^{i}B(i,j,1-f)\label{eqn:hyper-failure2}
\end{align}

\subsection{Tuning Parameters}{\label{tuning}}
\subsubsection{Security vs Overhead:}\label{sec-overhead-metric}
Our detection algorithm has three tunable parameters $(N,K,Th)$ (see Table \ref{tab1} for description). In this section
we introduce two evaluation metrics: \emph{security} ($\psi$) and \emph{overhead} ($\eta$). We then tune $K$ and
$\mathit{Th}$ in terms of these evaluation metrics. We define \emph{security} as the probability of not choosing a
compromised circuit for actual usage and \emph{overhead} as the expected number of probes required for each usable
circuit (by usable circuits we refer to the circuits that are actually used by a client). We define $\psi$ and $\eta$
using the following functions (both of these metrics are approximations):\nolinebreak
\begin{align}
\psi&=1-\frac{gt\times\Pr(FN)}{gt\times\Pr(FN)+(1-g)(1-t)^2\times(1-\Pr(FP))}\label{eqn:psi}\\
\eta&=\frac{1+\left[gt+(1-g)(1-t)^2\right]\times K}{gt\times\Pr(FN)+(1-g)(1-t)^2\times(1-\Pr(FP))}\label{eqn:eta}
\end{align}
Detailed derivations of the metrics are given in Appendix \ref{derivation}. Figure \ref{sec_vs_overhead}, shows the
distribution of $(K,\mathit{Th})$ for different values of $\psi$ and $\eta$. Since we are interested in attaining
higher security guarantee at a lower overhead, we cap the overhead axis to 25 probes per usable circuit. We can see
that as $K$ increases the overhead per usable circuit also increases which is expected. On the other hand an increase
in $\mathit{Th}$ (i.e., as it approaches $K$) results in an increase in $FP$ which causes overhead to rise. So
typically we want to choose a $(K,\mathit{Th})$ pair that achieves reasonable security guarantee and at the same time
induces acceptable overhead. Figure \ref{sec_vs_overhead} also highlights the fact that as $g$ increases the security
metric $\psi$ decreases. From figure \ref{sec_vs_overhead} we see that values such as $(K,\mathit{Th})=(3,2)$ achieves
acceptable outcome in terms of both security and overhead. We also discuss probabilistic bounding of our parameters in
Appendix \ref{pr-tunning} and show an alternative tuning process where we use the crossover point of $FN$ and $FP$ to
determine the value of $(K,\mathit{Th})$ pair in Appendix \ref{fn-fp-crossover}.

\begin{figure}[!h]
\begin{tabular}{cc}
\epsfig{file=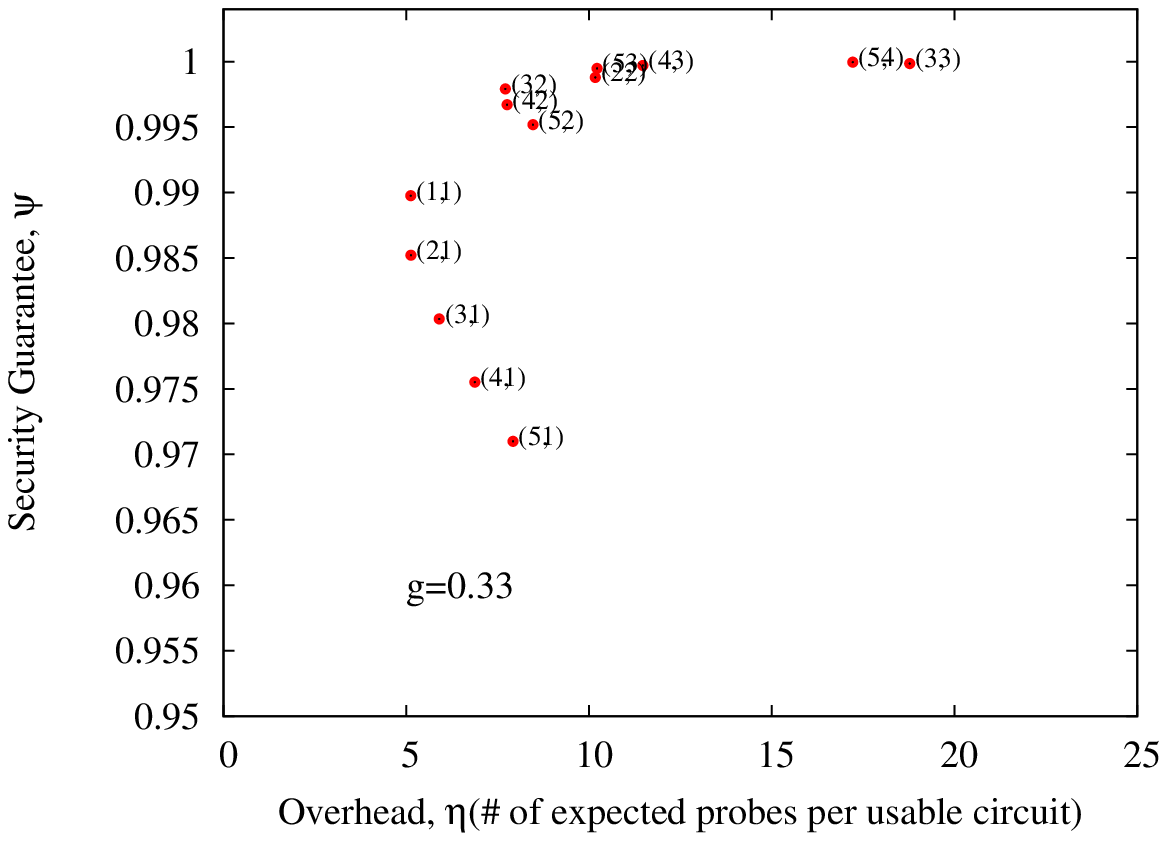,width=0.5\linewidth,clip=}&
\epsfig{file=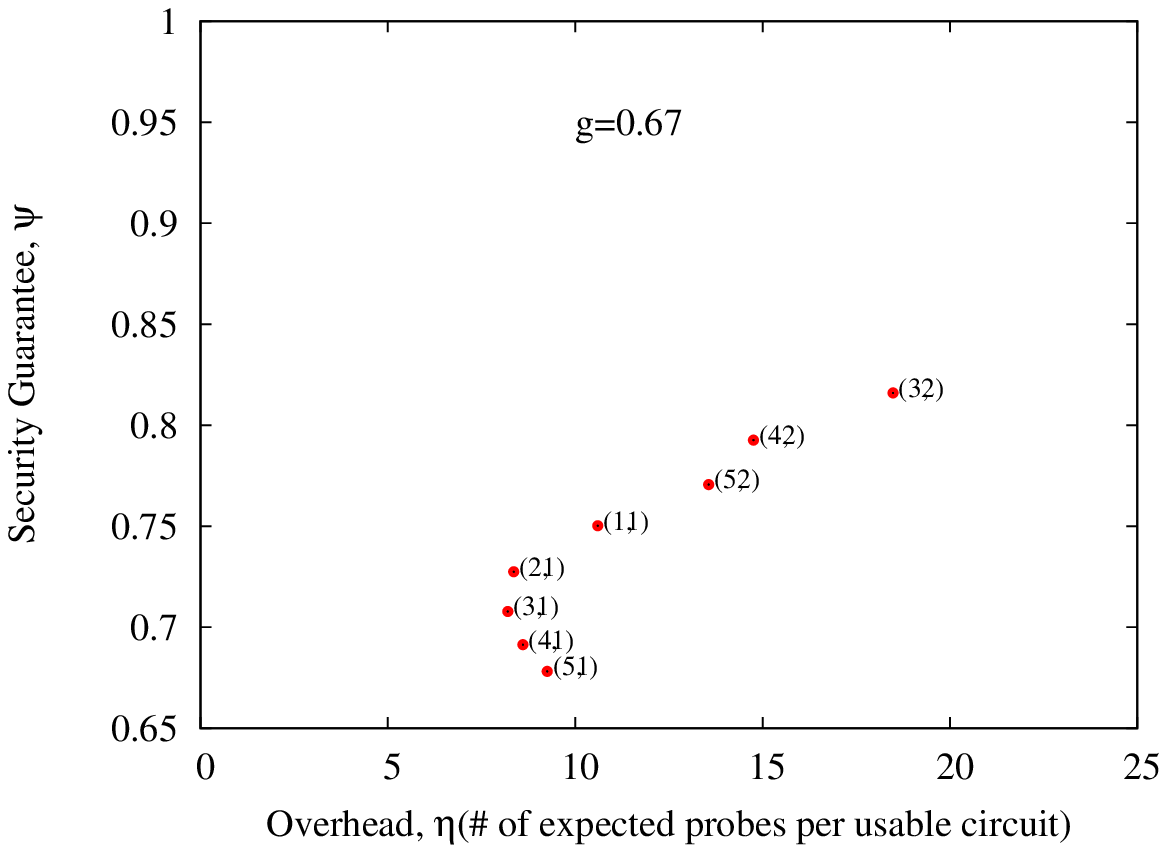,width=0.5\linewidth,clip=}
\end{tabular}
\caption{Tuning parameters ($K,Th$) against security metric $\psi$ and communication overhead $\eta$ for different
fraction of compromised guards (per user). Results for $g=1/3,2/3$ are more interesting than the results for $g=0,1$ as
they are trivial.} \label{sec_vs_overhead}
\end{figure}

\subsection{Analyzing Other Attack Strategies}{\label{strategy}}
So far we have assumed that the adversary is doing selective DoS, i.e., dropping any communication which it cannot
compromise. However, it is possible for the adversary to do probabilistic dropping, where a compromised router
terminates a certain fraction of all circuits that it cannot compromise. Doing so could potentially increase an
adversaries chance of passing the second phase. Lets analyze how probabilistic dropping affects our detection
algorithm. Any circuit formation must belong to the set $\{HHH,HHC,HCH,CHH,CCH,CHC,HCC,CCC\}$. Now circuits of forms
$\{HHH,CHC,CCC\}$ are never terminated (ignoring network failure), but under selective DoS all other forms of circuits
are always dropped. However, if the adversary was not to kill any circuit then all forms of circuits would survive the
first phase and a compromised circuit would have a wider variety of circuits to chose from in the second phase. Any
circuit belonging to the set $\{HHC,CHC,CCC,HCC\}$ would benefit a compromised circuit in the second phase. Under
selective DoS the probability of selecting another circuit with compromised exit router is
$\frac{gt}{gt+(1-g)(1-t)^2}$. On the other hand with $d\%$ drop rate this probability becomes:\nolinebreak
\begin{equation}
\frac{gt+[(1-g)(1-t)t+(1-g)t^2](1-d)}{(1-g)(1-t)^2+gt+[1-(1-g)(1-t)^2-gt](1-d)}\; \label{eqn:prcxcdrop}
\end{equation}
So a seemingly better strategy against our detection algorithm would be to allow all circuits belonging to the set
$\{HHC,CHC,CCC,HCC\}$ to always go through and drop other forms of circuits at rate $d$. We will call this strategy as
\textit{shrewd} strategy for future reference. For the \textit{shrewd} strategy the probability of selecting a
\textit{candidate} with compromised exit router becomes:$\frac{t}{(1-g)(1-t)^2+t+[1-(1-g)(1-t)^2-t](1-d)}$

\begin{figure}[!h]
\centering
\begin{tabular}{cc}
\epsfig{file=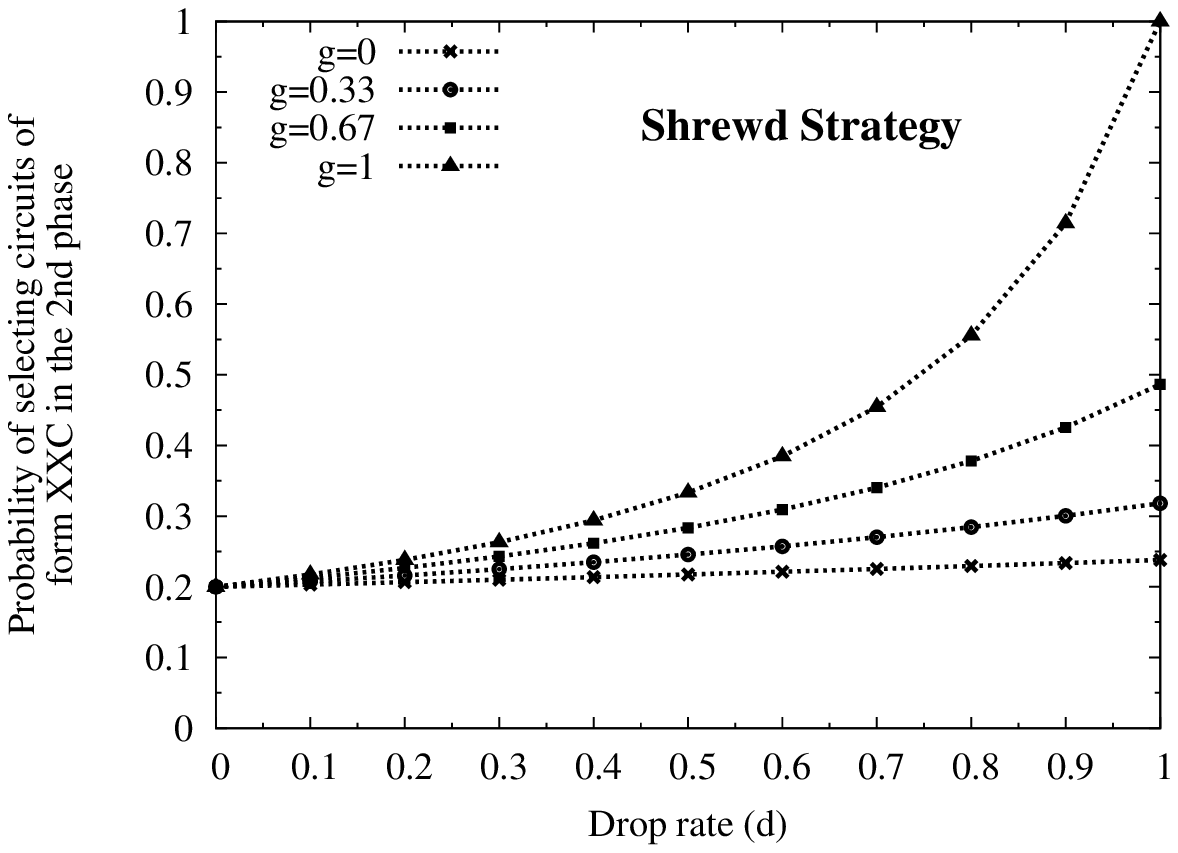,width=0.45\linewidth,clip=}&
\epsfig{file=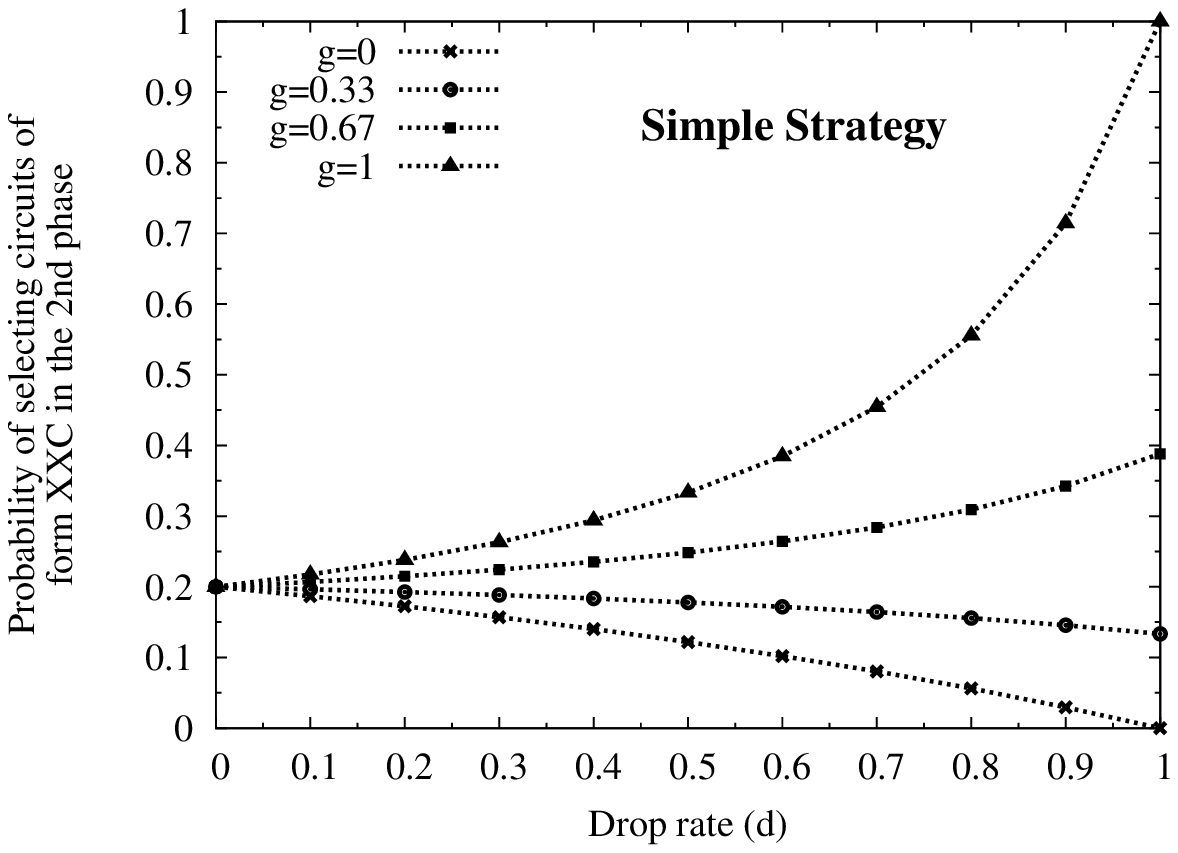,width=0.45\linewidth,clip=}
\end{tabular}
\caption{Probability of selecting candidate circuits (in the second phase) with compromised exit relay for different
drop rates $d$. Shrewd strategy seems to be a better choice for an adversary.} \label{shrewd-vs-simple}
\end{figure}

\begin{figure}[!h]
\centering
\begin{tabular}{cc}
\epsfig{file=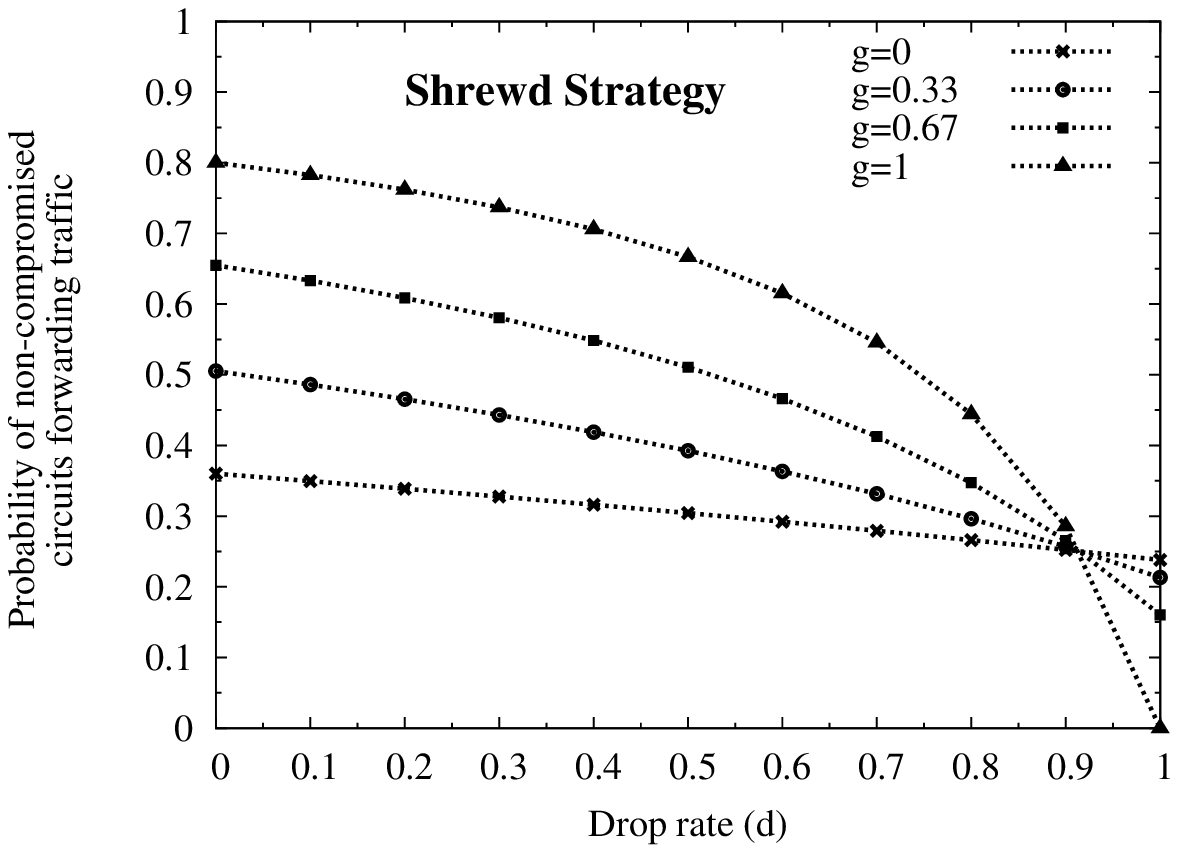,width=0.45\linewidth,clip=}&
\epsfig{file=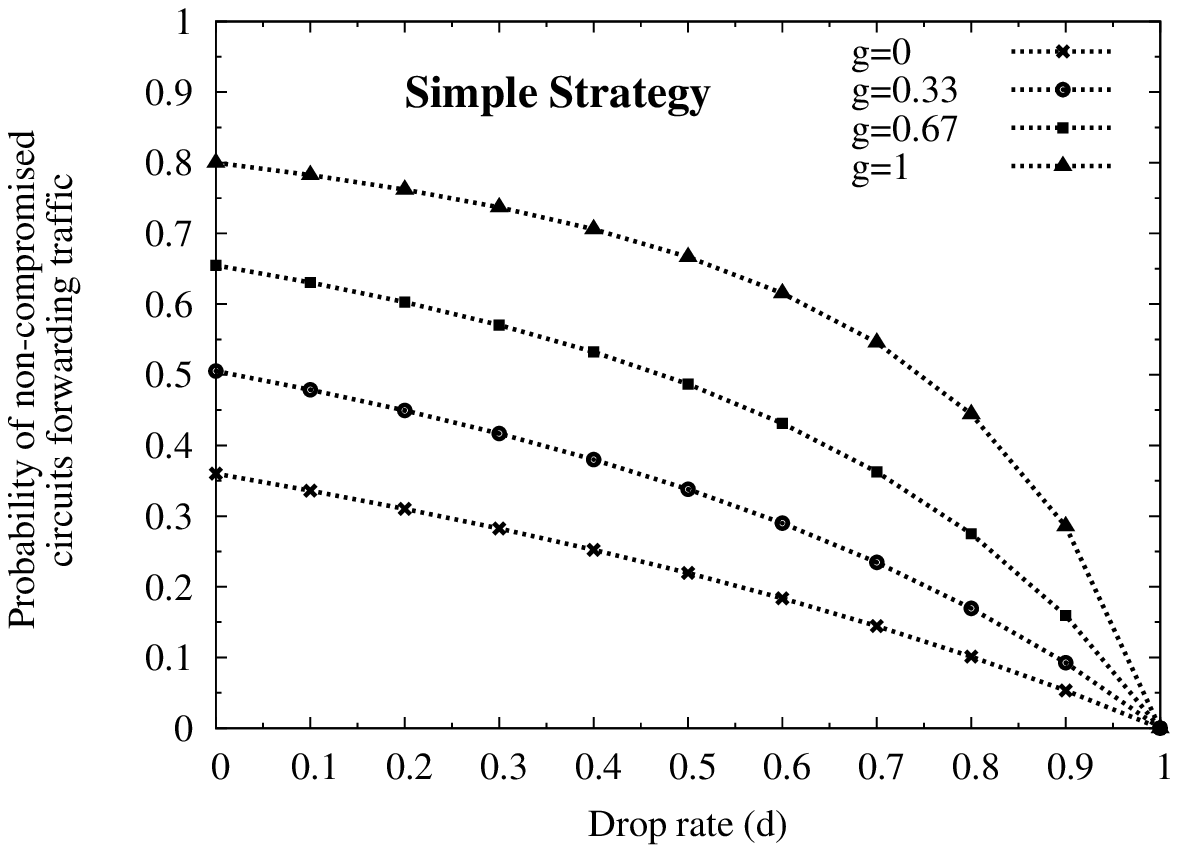,width=0.45\linewidth,clip=}
\end{tabular}
\caption{Fraction of non-compromised circuits that forward traffic under both strategies for different drops rate $d$.
Shrewd strategy seems to be more costly than the simple strategy.} \label{shrewd-vs-simple_cost}
\end{figure}

Figure \ref{shrewd-vs-simple} highlights the above calculated probability for different drop rates. As we can see from
figure \ref{shrewd-vs-simple}, at higher drop rate the probability of success in the second phase is significantly
higher for the \textit{shrewd} strategy compared to the simple strategy. Thus, with probabilistic dropping an adversary
can increase its chance of bypassing our detection algorithm; however that comes at the cost of forwarding traffic
through non-compromised circuits (circuits belonging to the set $\{HHC,HCH,CHH,CCH,HCC\}$ are termed as
non-compromised) which in turn can help honest circuits crossover the threshold more readily. We can calculated the
probability of forwarding traffic through non-compromised circuits using the following equation:\nolinebreak
\begin{equation}
\frac{\left[1-t-(1-g)(1-t)^2\right](1-d)+(1-g)t}{\left[1-t-(1-g)(1-t)^2\right](1-d)+t+(1-g)(1-t)^2}\;
\label{eqn:shrewd}
\end{equation}
Figure \ref{shrewd-vs-simple_cost} shows the fraction of non-compromised circuits used in forwarding traffic. We can
see from figure \ref{shrewd-vs-simple_cost} that \textit{shrewd} strategy forwards more traffic through non-compromised
circuits compared to the simple strategy. So the \textit{shrewd} strategy also helps honest circuits to crossover the
threshold, $\mathit{Th}$.

\subsection{Masking Probes}\label{masking-probes}
Note that it is important that compromised relays should not be able to distinguish between probes and real requests.
Otherwise the attacker can let probes go through and launch selective DoS on actual traffic. To mask probes from actual
user traffic we propose downloading popular web pages listed by Alexa \cite{popularwebsites}. 
This will make it harder for compromised exits to distinguish probes from actual traffic. Alternatively, we could
randomize our probes by performing automated random web searches \cite{trackmenot} and downloading the resulting pages.
Note that recent work shows that random web searches can be distinguished from real ones \cite{trackmetoo}; however,
this relies on building statistical profiles of search queries. In our case, every download is not linkable (at least
for circuits that the adversary can't compromise, which are the ones that matter!) and thus such techniques are not
applicable.

To eliminate any form of probing trace we can randomly choose $K$ \emph{non-repetitive middle nodes} along with the $K$
\textit{candidate} circuits. Doing so will have negligible impact on honest circuits because the probability of
selecting an honest middle node after first phase is: $\frac{(1-t)^3+t^2(1-t)}{(1-t)^3+t^2}$ (for $t=0.2$ this
probability is $\approx0.99$). Moreover, to remove temporal trace we can randomize the time at which we execute our
algorithm.

Now although we take different measures to make probes indistinguishable from real traffic. A strong adversary may come
up with some form of timing analysis to distinguish probes from real traffic. It should be noted that making probes
indistinguishable from real traffic is a hard research problem and a solution to this problem would be beneficial to
many other existing security enhancing schemes. For example Tor's \emph{Bandwidth-Measuring Directory Authorities}
\cite{bwauthority} probe relays to compare their actual observed bandwidth with their advertised bandwidth. So in this
case it is critical that probes are indistinguishable from user traffic, otherwise compromised relays can report false
bandwidth to these Directory Authorities and increase their chance of being selected during actual usage.


\section{Experimental Evaluation}{\label{evaluation}}

\subsection{Simulation Results}
We implemented a simulator in C++ that emulates the basic functionality of Tor circuit construction and selective DoS
attack. We collected real Tor node information from \cite{TorStatus} and randomly tagged 20\% ($t=0.2$) of the
bandwidth to be controlled by a compromised entity. We vary $g$ ($0\leq g\leq1$) and $d$ ($0\leq d\leq1$) to analyze
the robustness and effectiveness of our detection algorithm. Here, 100\% drop rate refers to selective DoS and 0\% drop
means no dropping at all. In the following evaluations we give more emphasis to $g=1/3,2/3$ (i.e., 1 or 2 of the 3
guards is/are compromised), since $g=0,1$ are trivial scenarios. We set $K=3$ and $\mathit{Th}=2$ in most of the
simulations. To approximate the failure rate present in the current Tor network we use the TorFlow project
\cite{torflow} and set $f=0.23$ in all our simulations (details are given in Appendix \ref{failure-rate}). All
simulation results are averaged over 100 runs (along with their 95\% confidence interval).

\subsubsection{Robustness:}
\begin{figure}[!h]
\centering
\begin{tabular}{cc}
\epsfig{file=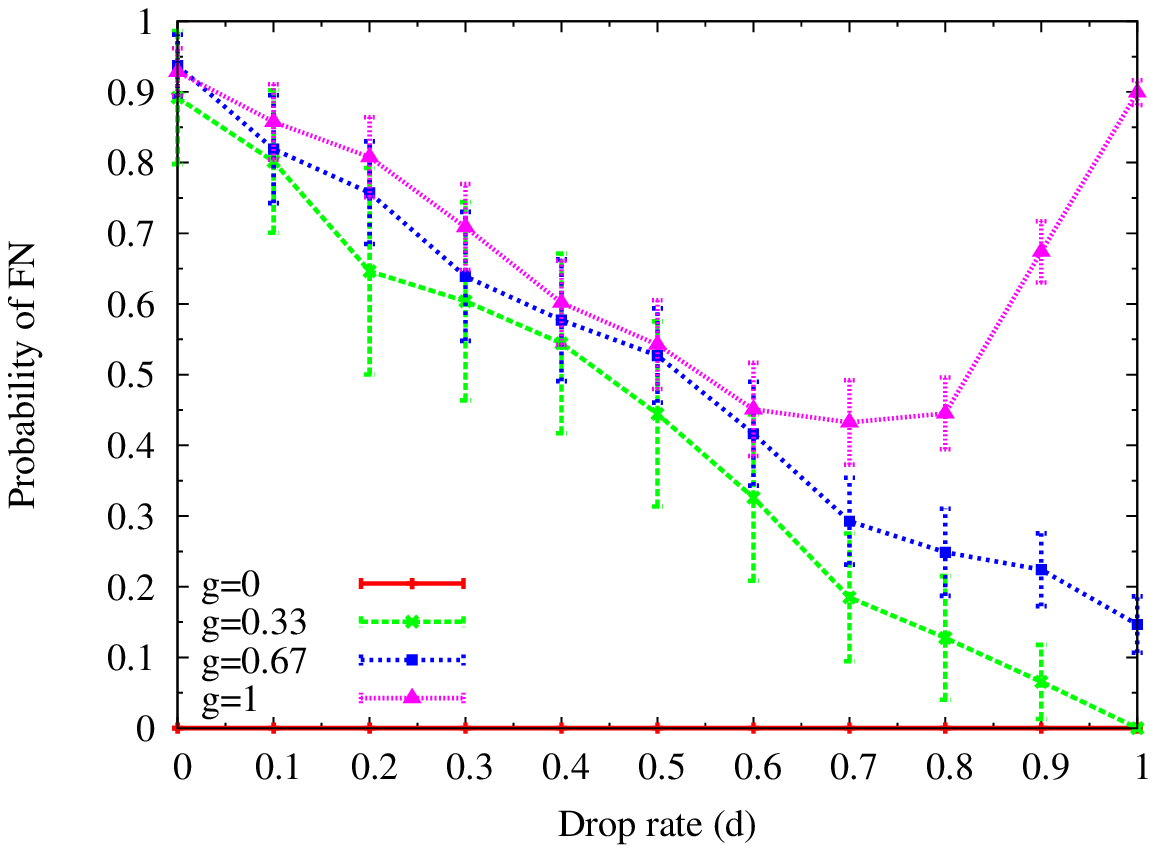,width=0.4\linewidth,clip=}&
\epsfig{file=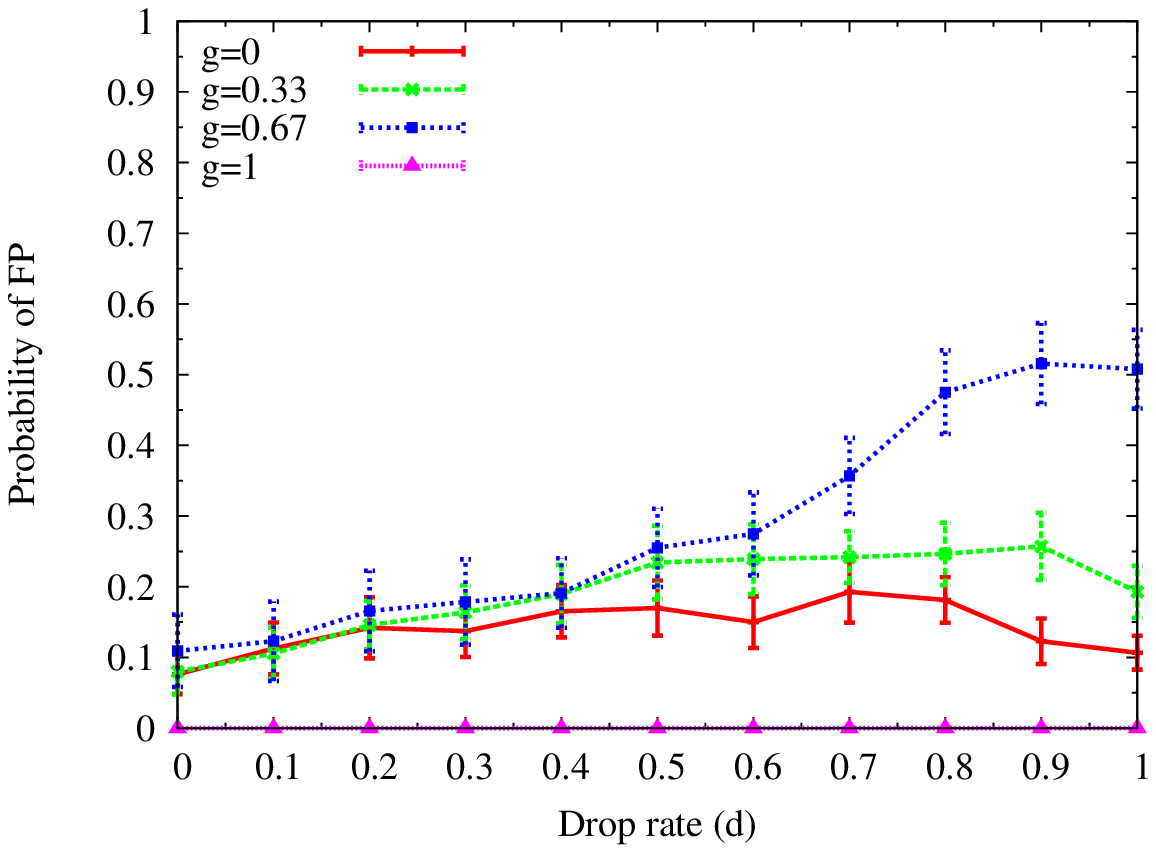,width=0.4\linewidth,clip=}
\end{tabular}
\caption{Probability of $FN$ and $FP$ for different drop rates $d$. In general, $FN$ decreases as $d$ rises while $FP$
increases as $d$ rises.} \label{FN-FP}
\end{figure}

First, we will look at the robustness of our detection algorithm in filtering out compromised circuits. For this
purpose we go through a series of evaluations. First, we investigate the $FN$ and $FP$ of our detection algorithm.
Figure \ref{FN-FP} shows the probability of $FN$ and $FP$ against different drop rates, $d$. From the figure we see
that as drop rate $d$ increases, $FN$ decreases for $0\leq g\leq2/3$. The main reason behind the decrease of $FN$ lies
on the fact that as compromised nodes start to perform aggressive dropping, the pool of available circuits in the first
phase quickly converges to the set $\{CXC,HHH\}$. This in turn lowers a compromised circuit's chance of selecting other
compromised candidate circuits in the second phase of our detection algorithm as honest circuits dominate over
compromised circuits for $t=0.2$. For $g=1$, $FN$ initially decreases until $d=0.6$; but for larger drop rates
($d>0.6$) $FN$ increases because with $g=1$ the set of available circuits in the first phase conform to the set
$\{CHC,CCC\}$ which increases a compromised circuit's chance of crossing the threshold of our detection algorithm. $FP$
tends to increase as the drop rate $d$ increases. The reason behind this is that as $d$ rises the probability of
selecting \textit{candidate} circuits of form $CXC$ increases which makes it harder for honest circuits to cross the
threshold, $\mathit{Th}$.

\begin{figure}[!h]
\centering
\begin{tabular}{cc}
\epsfig{file=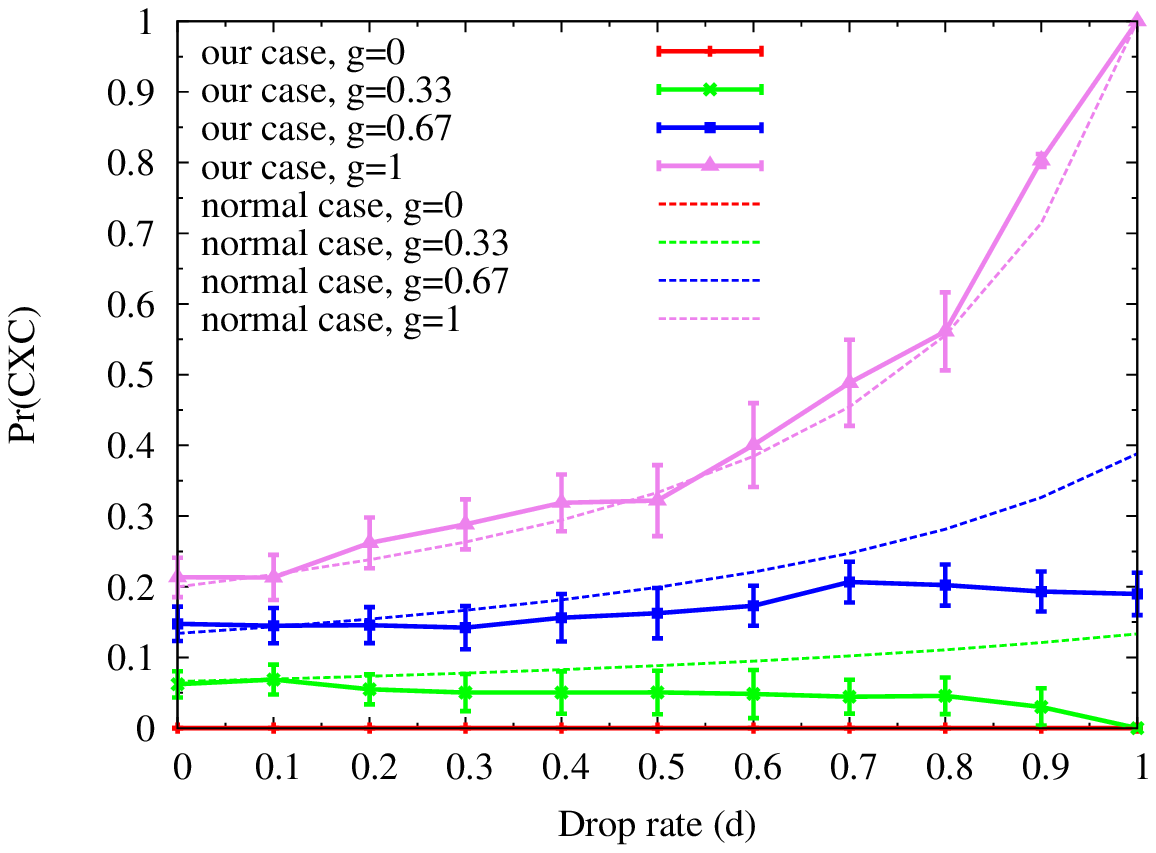,width=0.4\linewidth,clip=}&
\epsfig{file=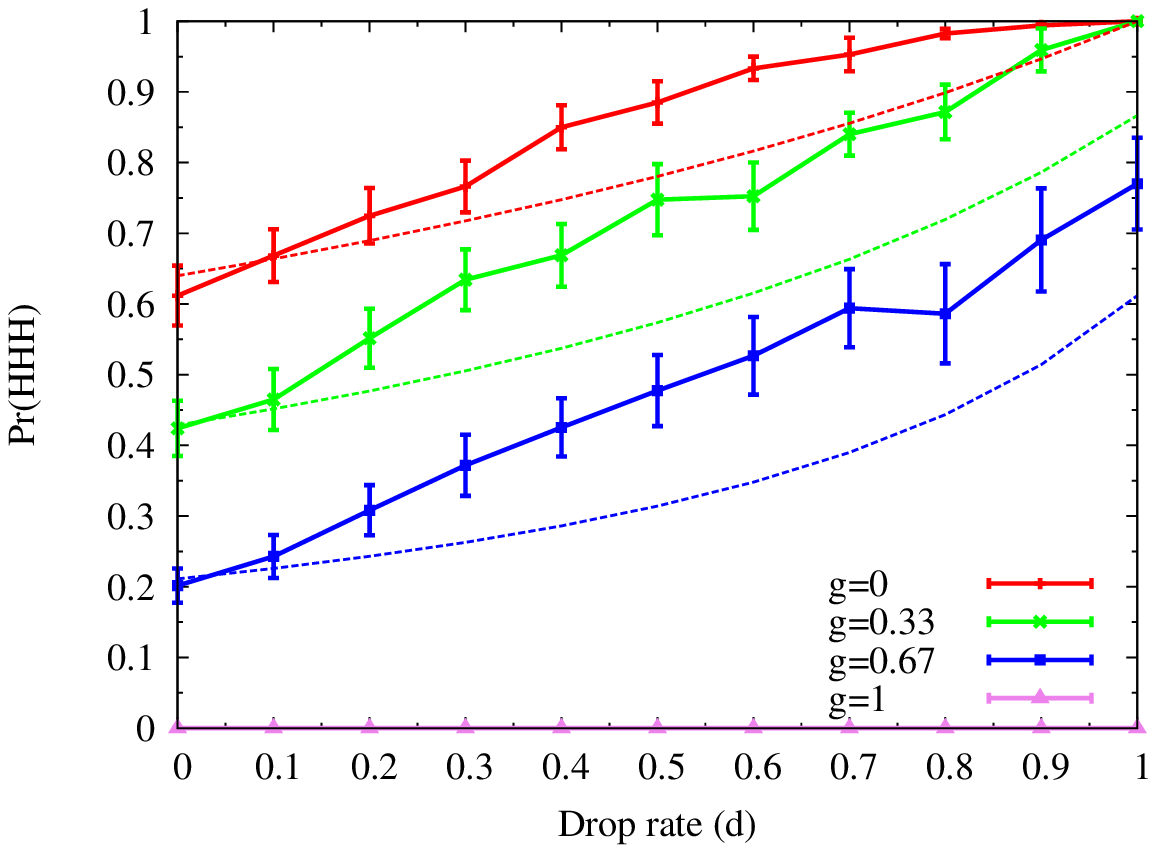,width=0.4\linewidth,clip=}
\end{tabular}
\caption{Probability of selecting compromised circuits $\Pr(CXC)$ and honest circuits $\Pr(HHH)$ for different drop
rates $d$. We see that $\Pr(CXC)$ decreases while $\Pr(HHH)$ increases as $d$ rises which signifies the robustness of
our detection algorithm against selective DoS attack.} \label{prob-cxc-hhh}
\end{figure}

Next, we measure the probabilities of selecting compromised and honest circuits once we have filtered the potential
compromised circuits. We compute these probabilities using the following equations:\nolinebreak
\begin{eqnarray}
    \Pr(CXC)=\frac{n(CXC)}{n(HHH)+n(CXC)+(1-d)n(Others)}\\
    \Pr(HHH)=\frac{n(HHH)}{n(HHH)+n(CXC)+(1-d)n(Others)}
\end{eqnarray}
where $\Pr(Others)$ represents the probability of selecting circuits other than $CXC$ or $HHH$. Figure
\ref{prob-cxc-hhh} shows the measured probabilities. We can see that $\Pr(CXC)$ decreases while $\Pr(HHH)$ increases as
drop rate $d$ increases. In the figure, we also highlight the corresponding probabilities for conventional Tor network
(indicated by the dashed lines). We see a significant improvement in filtering out compromised circuits compared to the
conventional Tor network (the only exception is when $g=1$ which is already a hopeless scenario). So from these results
we can say that our detection algorithm provides higher level of security assurance compared to conventional Tor.

\begin{figure}[!h]
\centering
\begin{tabular}{c|c}
\epsfig{file=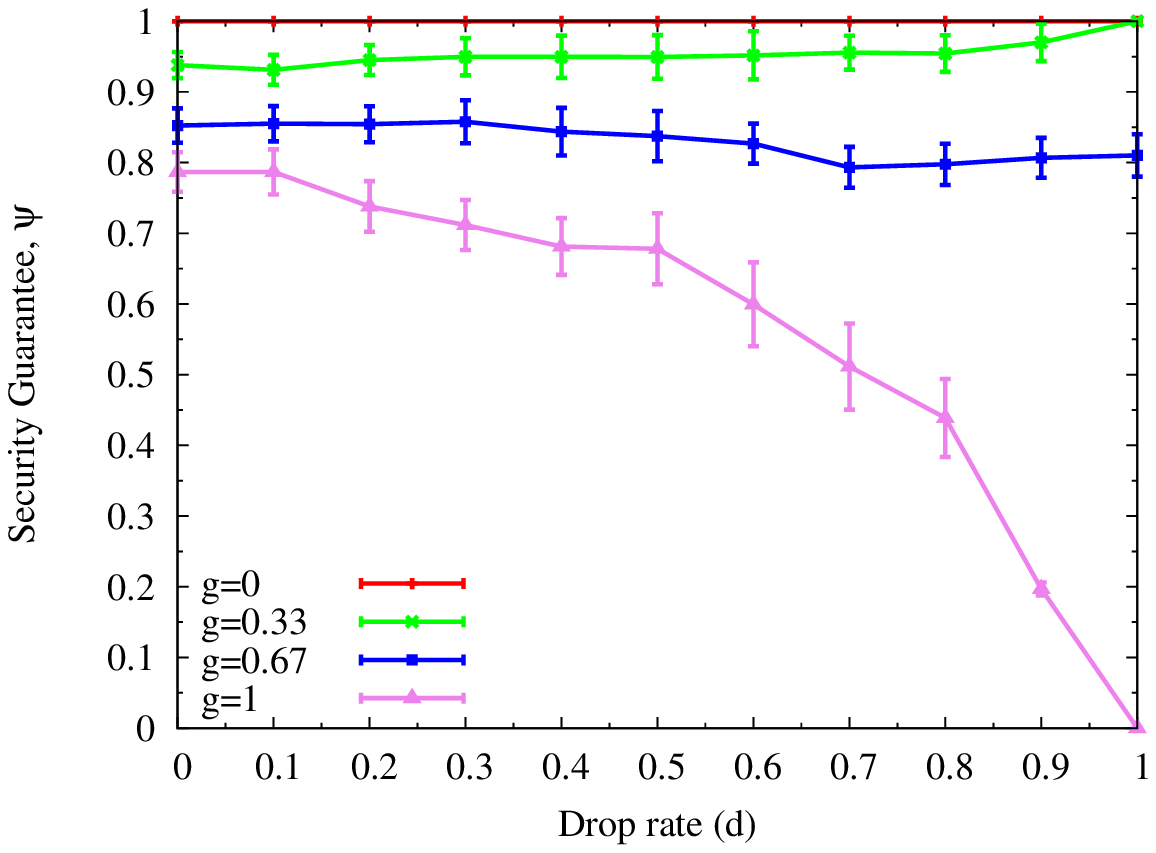,width=0.4\linewidth,clip=}&\epsfig{file=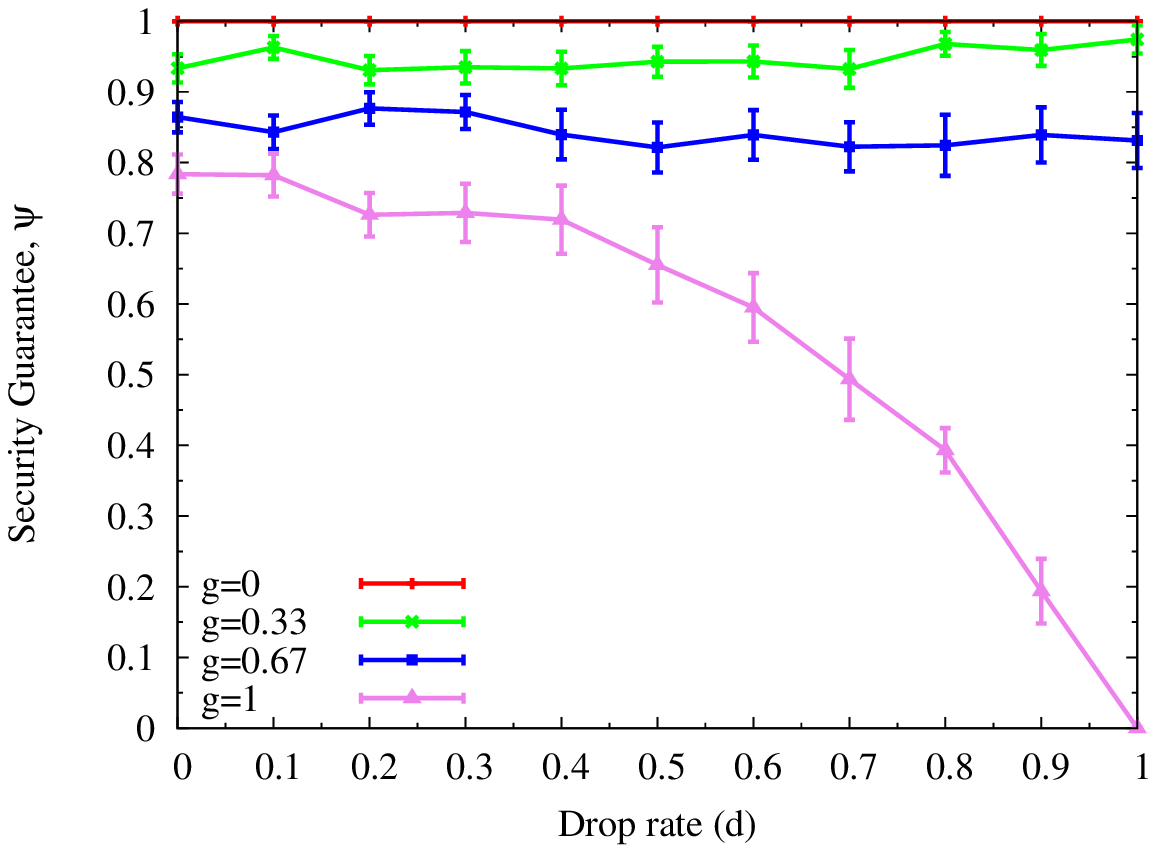,width=0.4\linewidth,clip=}\\
Simple Strategy& Shrewd Strategy
\end{tabular}
\caption{Comparing \textit{simple} strategy against \textit{shrewd} strategy for different probabilities.}
\label{shrewd-vs-simple-strategy}
\end{figure}

Now, we compare the \textit{shrewd} strategy (as described in Section \ref{strategy}) with the \textit{simple}
strategy. In section \ref{strategy} we saw that under \textit{shrewd} strategy $FN$ increases while $FP$ decreases as
drop rate $d$ increases.
To better understand the comparison between the two strategies we redefined the security metric defined in equation
(\ref{eqn:psi}) as:\nolinebreak
\begin{equation}
\psi=1-\frac{\Pr(CXC)}{\Pr(CXC)+\Pr(HHH)+(1-d)\Pr(Others)}
\end{equation}
This redefined metric will help us determine whether the change in $\Pr(CXC)$ dominates over the change in $\Pr(HHH)$
for the \textit{shrewd} strategy (as both probabilities increase for the \textit{shrewd} strategy). Figure
\ref{shrewd-vs-simple-strategy} plots the redefined metric $\psi$ against drop rate $d$. As we can see from the figure
both strategies have similar outcome. This means that the adversary really does not gain anything by adopting the
\textit{shrewd} strategy.

\subsubsection{Security vs Overhead Tradeoff:}
\begin{figure}[!h]
\centering
\begin{tabular}{ccc}
K=3,Th=2 & K=4,Th=3 & K=5,Th=4\\
\epsfig{file=security_vs_drop_K_3_Th_2.eps,width=0.33\linewidth,clip=}&\epsfig{file=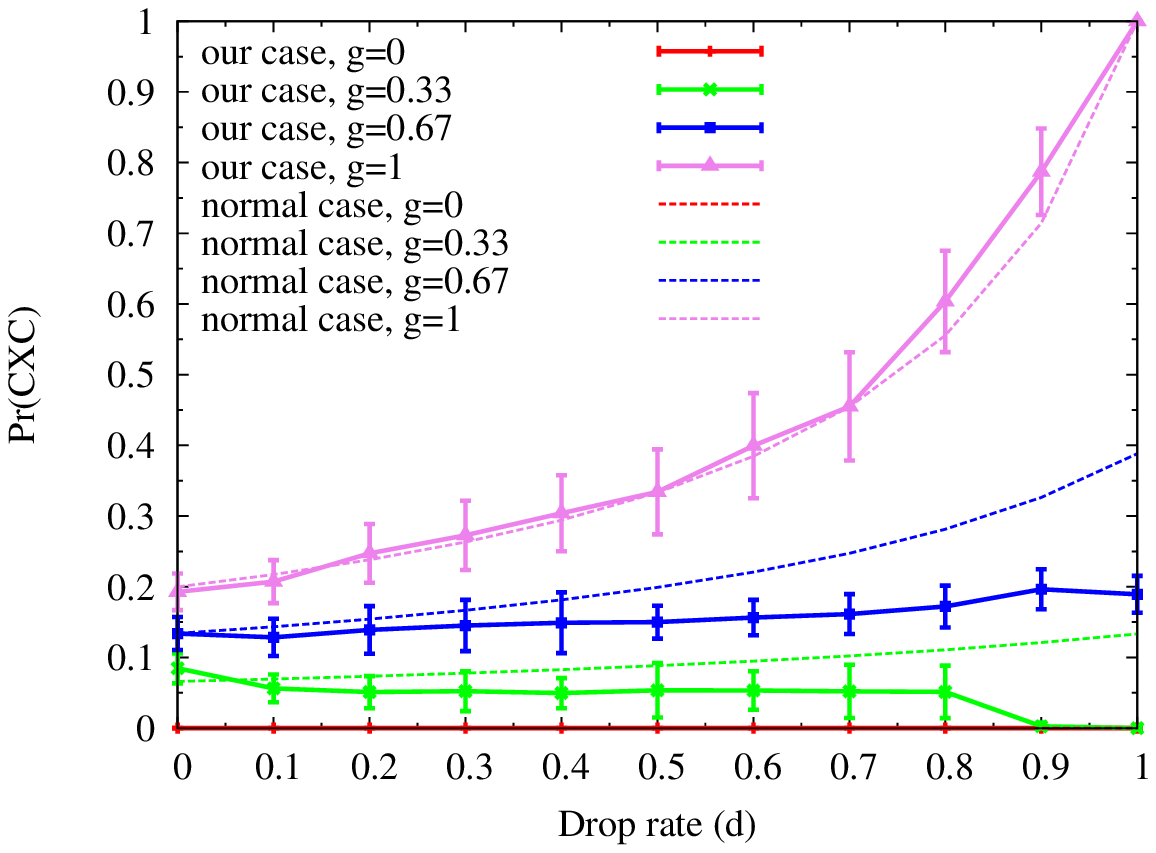,width=0.33\linewidth,clip=}&\epsfig{file=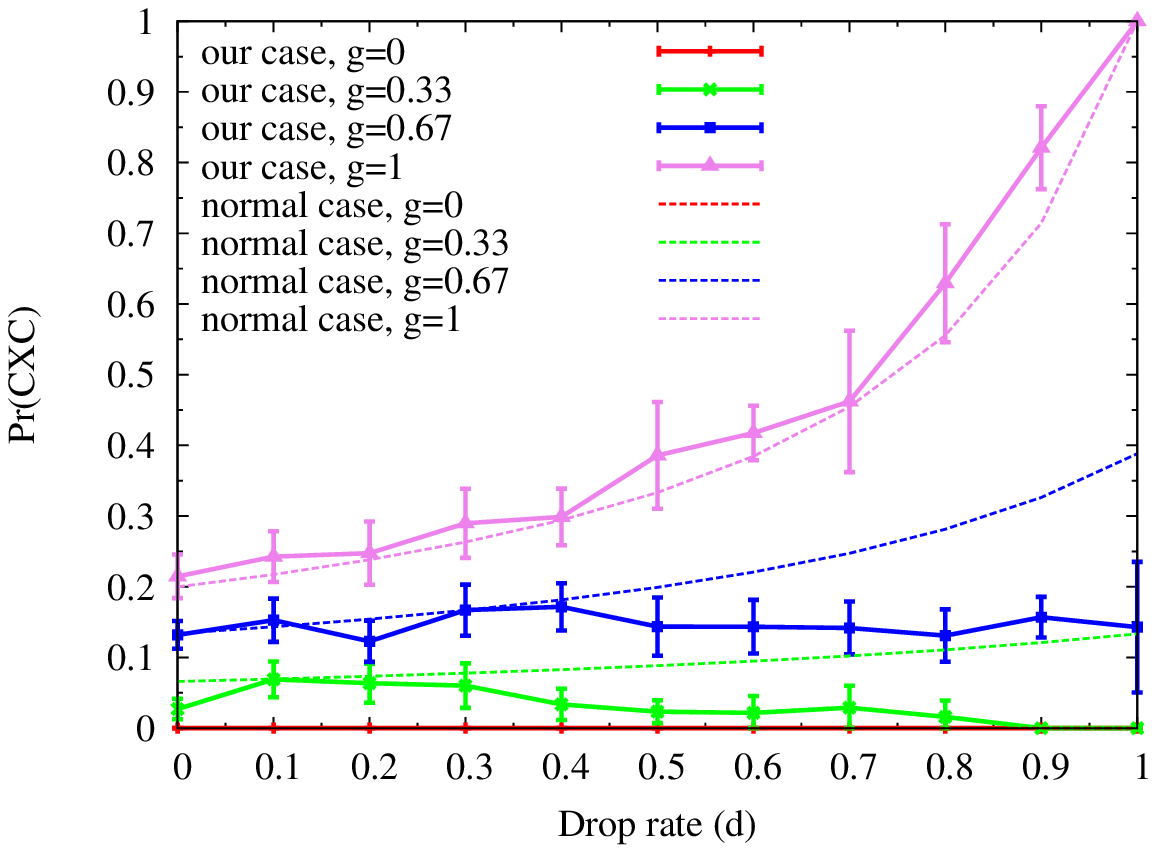,width=0.33\linewidth,clip=}\\
\epsfig{file=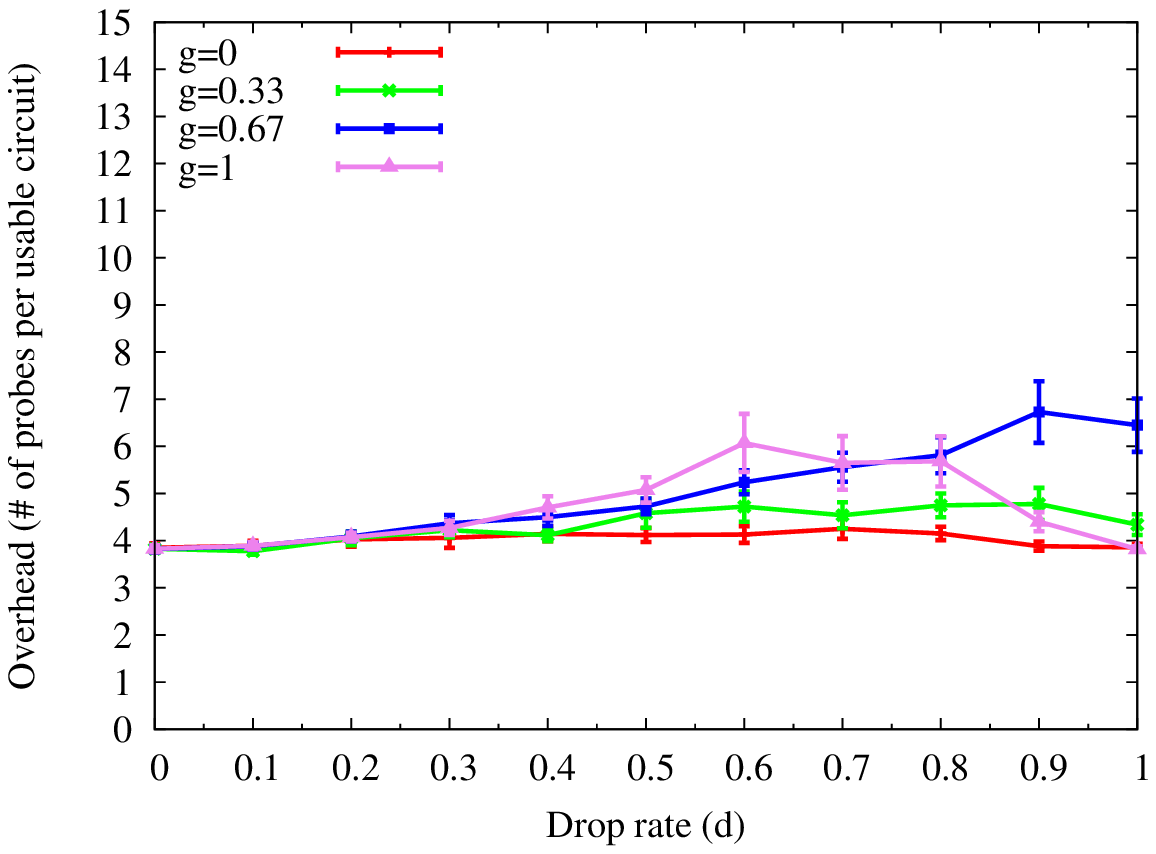,width=0.33\linewidth,clip=}&\epsfig{file=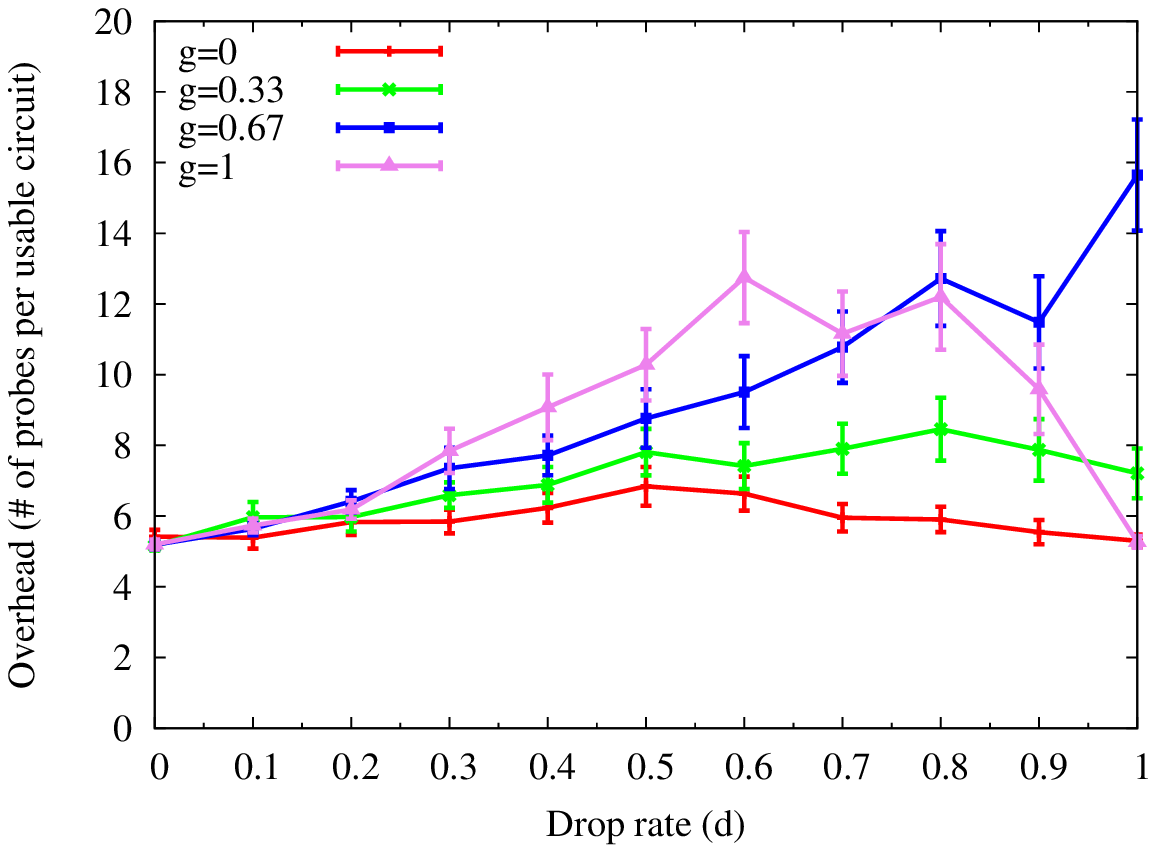,width=0.33\linewidth,clip=}&\epsfig{file=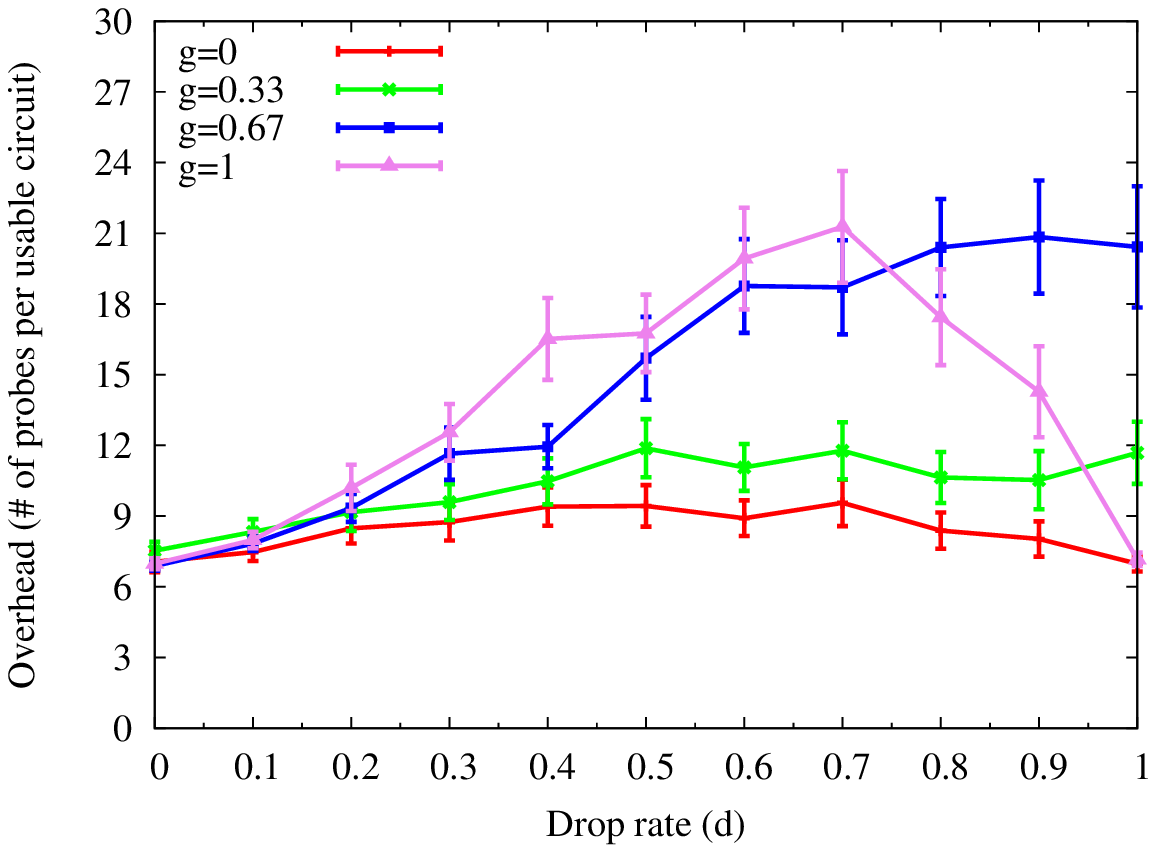,width=0.33\linewidth,clip=}
\end{tabular}
\caption{Average security guarantee(in terms of selecting a compromised circuits) and overhead per usable circuit for
different drop rates $d$. As we increase $\mathit{Th}$ the security guarantees improve but overhead per usable circuit
increases.} \label{varying-sec}
\end{figure}

In this section we explore the impact the pair $(K,\mathit{Th})$ has on the different metrics we have evaluated so far.
We evaluate the security guarantee (i.e., we calculate $\Pr(CXC)$) and overhead per usable circuit for different values
of $(K,\mathit{Th})$ pair. We found that the following effect dominates: lower values of $K$ improve overall overhead
(shown in figure \ref{varying-sec}) while higher values of $\mathit{Th}$ increase the security guarantee of the
detection algorithm (shown in figure \ref{varying-sec}). Thus, there exists a tradeoff between the obtainable security
guarantee and the required bandwidth overhead.

\subsection{Real World Experiment}
We carried out our experiment by introducing our own relays into the Tor network, all of which acted as compromised
nodes. For this purpose we used 11 Emulab \cite{emulab} machines, 10 of which were configured to act as Tor relays with
a minimum bandwidth capacity of 20Kbps. Note that all our nodes belonged to the same /16 subnet, meaning that no user
would (by default) choose two of our nodes in the same circuit. Moreover, to prevent other users from using our nodes
as exit nodes, we configured our relays with a fixed exit policy (allowing connection to only specific destinations).
All these measures were taken to reduce the non-benevolent impact of our relays on the Tor network.

For implementing selective DoS we take an approach similar to the one described by Bauer et al. \cite{Bauer2007}. Here,
out of the 11 machines we run Tor protocol on 10 of them and used the remaining machine as a server for gathering
timing information about which router is communicating with whom at what time. The sever performs timing analysis and
informs the other 10 machines when to drop communication to perform selective DoS (we modified Tor source code
version-\textit{tor-0.2.2.35}). We implemented our detection algorithm in the client side in Python (we used the open
source Python library of Tor Control Port (\textit{TorCtl.py})\cite{TorController}).

\subsubsection{Robustness:}
In our experiments we first query the Tor directory server to retrieve a list of all available Tor routers and then
filter this list by considering only those routers which are flagged as \textit{running}, \textit{stable} and
\textit{valid}, because we want our routers to be alive and running during our experiments. We selected 40 Tor nodes (3
guards, 19 exits and 18 relays) at random with probability proportional to their bandwidth and added our own 10 nodes
to this set to get a total of 50 nodes. Then we ran our experiments on this small set of Tor nodes where nodes are
selected randomly. This choice results in about 20\% of the nodes being compromised. To emulate user traffic/usage, we
retrieve random web files 100--300\,KB in size. We set $K=3,\mathit{Th}=2$ for our experiments. Table~\ref{real-exp}
summarizes our findings.

\begin{table}
\centering \caption{Experimental results from the real Tor network} \resizebox{5cm}{!}{
\begin{tabular}{|c|c|c|c|c|}
\hline
\multirow{2}{*}{$g$}&\multirow{2}{*}{$FN$}&\multirow{2}{*}{$FP$}&\multirow{2}{*}{$\psi$}&Security in\\
&&&&Current Tor\\
\hline
0&0.0&0.0664&1.0&1.0\\
\hline
1/3&0.0&0.178&1.0&0.867\\
\hline
2/3&0.133&0.283&0.843&0.612\\
\hline
1&1.0&0.0&0.0&0.0\\
\hline
\end{tabular}}
\label{real-exp}
\end{table}
From table \ref{real-exp} we see that as $g$ increases the security assurance provided by both our approach and the
conventional Tor network go down. However, for $g=1/3,2/3$ our approach shows significant improvement in filtering out
compromised circuits. Thus we can say that our detection algorithm provides strong robustness even when two guards out
of the three guard nodes are compromised.

\subsubsection{Overhead:}
Let us now estimate what kind of bandwidth overhead our mechanism would inflict on the real Tor network. From figure
\ref{varying-sec} we see that on average a single usable circuit would require approximately 4 probes. Since we are
proposing to use random popular web sites as probing destinations we can approximate the average probe size to be 300KB
\cite{webmetric}. Moreover, since users requires 6 usable circuits every one hour we set the time interval to $3600$
seconds. From these values we calculate the bandwidth required per usable circuit to be
$\frac{3\times300\times4}{3600}=1$KB/s. So for 6 usable circuits a user would consume 6KB/s of bandwidth. Now, Tor's
bandwidth capacity was found to be 3,359,120 KB/s \cite{torbandwidth} during the month of September, 2012. If we allow
5\% of the bandwidth to be used for our detection algorithm then we can support approximately 28,000
\textit{simultaneous} users per hour (i.e.,$\approx672,000$ users daily which is $>619,696$, the peak daily Tor users
for Oct, 2012 \cite{torbandwidth}). By \textit{simultaneous} users we refer to users that run our algorithm exactly at
the same time. However, we could lower our bandwidth requirement by lowering the individual probe size (potentially we
could use smaller size popular web pages as probe destinations). We could also use a client's circuit usage history to
construct the initial set of working circuits (in our model this is represented by $N$). Alternatively, we could spread
out our filtering scheme over a longer time frame.

\section{Related Work}{\label{related_work}}
Borisov et al. \cite{Borisov2007} first showed that carrying out selective DoS could benefit an adversary to increase
its chance of compromising anonymity for both high and low-latency anonymous communication systems like Hydra-Onion
\cite{hydra-onion}, Cashmere \cite{cashmere} and Salsa \cite{salsa}. In fact, it was pointed out that with 20\%
compromised nodes in Salsa, the selective DoS attack results in 19.14\% compromised tunnels compared to the
conventional security analysis of 6.82\% compromised tunnels.

Later on Danner et al. \cite{Danner2009} proposed a detection algorithm for selective DoS attack on Tor. Their
algorithm basically probes each individual Tor node in the network and they prove that this requires $O(n)$ probes to
detect all compromised nodes in the Tor network comprising of $n$ participants. For Tor circuits of length 3 their
algorithm requires $3n$ probes; however to handle transient network failures they proposed to repeat each probe $l$
number of times. They define an lower bound of $l$; under conventional analysis a value of 10 for $l$ was shown to be
sufficient. So this means for a Tor network with 3000 nodes, it would require roughly $90,000$ probes to identify all
compromised nodes in the network. However, their algorithm assumes that compromised nodes have a fixed characteristic
of always dropping non-compromised circuits. They do not consider complex attack strategies whether compromised nodes
may perform random dropping. Moreover, they provide no analysis of the overhead involved in performing these probes.
Clearly, this approach will not scale well if a large number of users start to probe Tor nodes simultaneously. We take
a different approach where a user tries to accumulate a set of non-compromised working circuits (for future usage)
instead of classifying all Tor nodes as either compromised or non-compromised. We also analyze different types of
complex attack strategies against our detection algorithm.

Recently (on 16 Oct, 2012), Mike Perry (Torbutton and Tor Performance Developer) proposed a client-side accounting
mechanism that tracks the circuit failure rate for each of the client's guards~\cite{torproposal209}. The goal is to
avoid malicious guard nodes that deliberately fail circuits extending to non-colluding exit nodes. We take a more
proactive approach to finding malicious circuits through probing instead of tracking actual circuit usage.

\section{Conclusion}{\label{conclusion}}
Anonymous communication systems like Tor are vulnerable to selective denial of service attack which considerably lowers
anonymity. Such attacks however, can be detected through probing. Our detection algorithm probes communication channels
to filter out potentially compromised ones with high probability. We also show that adaptive adversaries who choose to
deny service probabilistically do not benefit from adopting such strategy. Our experimental results demonstrate that
our detection algorithm can effectively defend users against selective DoS attack.

\bibliographystyle{plain}
\bibliography{ourbib}

\begin{thebibliography}{10}

\bibitem{bwauthority}
Bandwidth scanner specification.
\newblock
  \url{https://gitweb.torproject.org/torflow.git/blob/HEAD:/NetworkScanners/BwAuthority/README.spec.txt}.

\bibitem{dutch}
Dutch government proposes cyberattacks against... everyone.
\newblock
  \url{https://www.eff.org/deeplinks/2012/10/dutch-government-proposes-cyberattacks-against-everyone}.

\bibitem{emulab}
Emulab.
\newblock \url{https://www.emulab.net}.

\bibitem{Freenet}
Freenet.
\newblock \url{https://freenetproject.org/}.

\bibitem{I2P}
I2p.
\newblock \url{http://www.i2p2.de/}.

\bibitem{popularwebsites}
Top sites on the web.
\newblock \url{http://www.alexa.com/topsites}.

\bibitem{TorController}
Tor controller.
\newblock \url{https://svn.torproject.org/svn/blossom/trunk/TorCtl.py}.

\bibitem{torbandwidth}
Tor metrics portal:.
\newblock \url{https://metrics.torproject.org/}.

\bibitem{torproposal209}
Tor proposal 209:.
\newblock
  \url{https://gitweb.torproject.org/user/mikeperry/torspec.git/blob/path-bias-tuning:/proposals/209-path-bias-tuning.txt}.

\bibitem{torflow}
Torflow project.
\newblock \url{https://gitweb.torproject.org/torflow.git}.

\bibitem{TorStatus}
Torstatus.
\newblock \url{http://torstatus.blutmagie.de/index.php}.

\bibitem{Bauer10onthe}
Kevin Bauer, Joshua Juen, Nikita Borisov, Dirk Grunwald, Douglas Sicker, and
  Damon Mccoy.
\newblock On the optimal path length for tor, 2010.

\bibitem{Bauer2007}
Kevin Bauer, Damon McCoy, Dirk Grunwald, Tadayoshi Kohno, and Douglas Sicker.
\newblock Low-resource routing attacks against {Tor}.
\newblock In {\em Proceedings of the 2007 ACM workshop on Privacy in Electronic
  Society}, WPES '07, pages 11--20, 2007.

\bibitem{Borisov2007}
Nikita Borisov, George Danezis, Prateek Mittal, and Parisa Tabriz.
\newblock Denial of service or denial of security?
\newblock In {\em Proceedings of the 14th ACM conference on Computer and
  communications security}, CCS '07, pages 92--102. ACM, 2007.

\bibitem{Chaum}
David~L. Chaum.
\newblock Untraceable electronic mail, return addresses, and digital
  pseudonyms.
\newblock {\em Commun. ACM}, 24:84--90, February 1981.

\bibitem{Danner2009}
Norman Danner, Danny Krizanc, and Marc Liberatore.
\newblock Detecting denial of service attacks in {Tor}.
\newblock In Roger Dingledine and Philippe Golle, editors, {\em Financial
  Cryptography and Data Security}, volume 5628 of {\em Lecture Notes in
  Computer Science}, pages 273--284. Springer Berlin / Heidelberg, 2009.

\bibitem{TorPathSpec}
Roger Dingledine and Nick Mathewson.
\newblock {Tor} path speciﬁcation.
\newblock
  \url{https://gitweb.torproject.org/torspec.git/blob/HEAD:/path-spec.txt}.

\bibitem{Dingledine2004}
Roger Dingledine, Nick Mathewson, and Paul Syverson.
\newblock Tor: The second-generation onion router.
\newblock In {\em Proceedings of the 13th conference on USENIX Security
  Symposium}, SSYM'04, pages 303 -- 320, 2004.

\bibitem{Hahn2010}
Sebastian Hahn and Karsten Loesing.
\newblock Privacy-preserving ways to estimate the number of {Tor} users,
  November 2010.
\newblock
  \url{https://metrics.torproject.org/papers/countingusers-2010-11-30.pdf}.

\bibitem{trackmenot}
Daniel~C. Howe and Helen Nissenbaum.
\newblock In Ian Kerr, Valerie Steeves, and Carole Lucock, editors, {\em
  Lessons from the Identity Trail: Anonymity, Privacy, and Identity in a
  Networked Society}, pages 417--436. Oxford University Press, 2009.

\bibitem{hydra-onion}
Jan Iwanik, Marek Klonowski, and Miroslaw Kutylowski.
\newblock Duo{\textendash}onions and hydra{\textendash}onions {\textendash}
  failure and adversary resistant onion protocols.
\newblock In {\em Proceedings of the IFIP TC-6 TC-11 Conference on
  Communications and Multimedia Security}, pages 1--15. Springer Boston,
  September 2004.

\bibitem{Neil2004}
Brian~Neil Levine, Michael~K. Reiter, Chenxi Wang, and Matthew Wright.
\newblock Timing attacks in low-latency mix systems.
\newblock In {\em Financial Cryptography}, pages 251--265. Springer, 2004.

\bibitem{Loesing2009}
Karsten Loesing.
\newblock Measuring the tor network: Evaluation of client requests to the
  directories.
\newblock Technical report, June 2009.
\newblock
  \url{https://metrics.torproject.org/papers/directory-requests-2009-06-25.pdf}.

\bibitem{salsa}
Arjun Nambiar and Matthew Wright.
\newblock Salsa: a structured approach to large-scale anonymity.
\newblock In {\em Proceedings of the 13th ACM conference on Computer and
  communications security}, CCS '06, pages 17--26, 2006.

\bibitem{overlier-syverson:oakland06}
Lasse Overlier and Paul Syverson.
\newblock Locating hidden servers.
\newblock In {\em Proceedings of the 2006 IEEE Symposium on Security and
  Privacy}, pages 100--114, 2006.

\bibitem{trackmetoo}
Sai Peddinti and Nitesh Saxena.
\newblock On the privacy of web search based on query obfuscation: A case study
  of trackmenot.
\newblock In {\em Privacy Enhancing Technologies}, volume 6205 of {\em Lecture
  Notes in Computer Science}, pages 19--37. Springer Berlin / Heidelberg, 2010.

\bibitem{Reed1998}
M.G. Reed, P.F. Syverson, and D.M. Goldschlag.
\newblock Anonymous connections and onion routing.
\newblock {\em IEEE Journal on Selected Areas in Communications}, 16(4):482 --
  494, May 1998.

\bibitem{Shmatikov2006}
Vitaly Shmatikov and Ming-Hsiu Wang.
\newblock Timing analysis in low-latency mix networks: Attacks and defenses.
\newblock In {\em Proceedings of ESORICS}, pages 18--33, 2006.

\bibitem{webmetric}
Google Sreeram~Ramachandran.
\newblock Web metrics: Size and number of resources.
\newblock \url{https://developers.google.com/speed/articles/web-metrics}.

\bibitem{Syverson2001}
Paul Syverson, Gene Tsudik, Michael Reed, and Carl Landwehr.
\newblock Towards an analysis of onion routing security.
\newblock In {\em International Workshop on Designing Privacy Enhancing
  Technologies: Design Issues in Anonymity and Unobservability}, pages 96--114.
  Springer-Verlag New York, Inc., 2001.

\bibitem{Wright2003}
Matthew Wright, Micah Adler, Brian~N. Levine, and Clay Shields.
\newblock Defending anonymous communications against passive logging attacks.
\newblock In {\em Proceedings of the 2003 IEEE Symposium on Security and
  Privacy}, SP '03, pages 28--41. IEEE Computer Society, 2003.

\bibitem{Wright2002}
Matthew~K. Wright, Micah Adler, Brian~Neil Levine, and Clay Shields.
\newblock An analysis of the degradation of anonymous protocols.
\newblock In {\em Proceedings of the Network and Distributed Security Symposium
  - NDSS {\textquoteright}02}. IEEE, February 2002.

\bibitem{Zhu2004}
Ye~Zhu, Xinwen Fu, Bryan Graham, Riccardo Bettati, and Wei Zhao.
\newblock On flow correlation attacks and countermeasures in mix networks.
\newblock In {\em Proceedings of Privacy Enhancing Technologies Workshop},
  pages 207--225, 2004.

\bibitem{cashmere}
Li~Zhuang, Feng Zhou, Ben~Y. Zhao, and Antony Rowstron.
\newblock Cashmere: resilient anonymous routing.
\newblock In {\em Proceedings of the 2nd conference on Symposium on Networked
  Systems Design \& Implementation - Volume 2}, NSDI'05, pages 301--314, 2005.

\end{thebibliography}

\normalsize
\section*{Appendix}
\appendix

\section{Approximating Failure Rate in the Tor Network}{\label{failure-rate}}
To approximate the failure rate present in the current Tor network we use the TorFlow
project \cite{torflow}. TorFlow project measures the performance of Tor network by creating Tor circuits and recording
statistical data such as circuit construction time, circuit failure rates and stream failure rate. We are interested in
the circuit failure rate as it directly impacts the false ratings of our detection algorithm. So for our purpose we run
the \emph{buildtime.py} \cite{torflow} python script to generate 10,000 Tor circuits and record their failure rate. We
ran the script 10 times and found the average failure rate to be approximately 23\%. We therefore set $f=0.23$ in our
simulations.

\section{Security and Overhead Metric}\label{derivation}
In this section we will show the step by step derivation of the security and overhead metrics defined in Section
\ref{sec-overhead-metric}. Now we defined our security metric $\psi$ as the probability of not selecting a compromised
circuit after running our filtering algorithm. Lets define $n(CXC)$ and $n(HHH)$ as the number of compromised and
honest circuits passing the first phase of our detection algorithm.\nolinebreak
\begin{align}
n(CXC)=N\times \frac{gt}{gt+(1-g)(1-t)^2},\hspace{1pt} n(HHH)=N\times \frac{(1-g)(1-t)^2}{gt+(1-g)(1-t)^2}\nonumber
\end{align}
Given the number of compromised and honest circuits that pass the first phase we can compute their corresponding
fractions that pass the second phase using our false negative and false positive rates. We can then derive $\psi$ using
the following equation:
\begin{align}
\psi&=1-\frac{n(CXC)\times\Pr(FN)}{n(CXC)\times\Pr(FN)+n(HHH)\times\left[1-\Pr(FP)\right]}\nonumber\\
&=1-\frac{gt\times\Pr(FN)}{gt\times\Pr(FN)+(1-g)(1-t)^2\times\left[1-\Pr(FP)\right]}
\end{align}

To compute the overhead per usable circuit we first need to
determine the total number of probes ($n(Probes)$) required by our
detection algorithm. We then compute the overhead metric $\eta$ by
dividing $n(Probes)$ by the total number of usable circuits.
\begin{align}
&n(Probes)=\frac{N}{{gt+(1-g)(1-t)^2}}+N\times K\nonumber\\
&\eta=\frac{n(Probes)}{n(CXC)\times\Pr(FN)+n(HHH)\times\left[1-\Pr(FP)\right]}\nonumber\\
&  =\frac{1+\left[gt+(1-g)(1-t)^2\right]\times K}{gt\times\Pr(FN)+(1-g)(1-t)^2\times\left[1-\Pr(FP)\right]}
\end{align}

\section{Probabilistic Bounding of Parameters}{\label{pr-tunning}}
We will use probabilistic expectation analysis to determine what values (or ranges) to use for them. By default, Tor
creates a new circuit every 10 minutes and since our algorithm is rerun every one hour, 6 honest circuits are required
by a user in an one hour period. So we can calculate the value of $N$ from this condition using the following function
$N=\left\lceil{6\times\frac{gt+(1-g)(1-t)^2}{(1-g)(1-t)^2}}\right\rceil$ So $N$ varies as $g$ varies. For $g=1$, $N$
tends to infinity which is understandable because with all guards being compromised a user can never construct an
honest circuit. In other words, if all the guards are compromised it's pointless to use any filtering technique. Now,
since a user does not know what fraction of its guards are compromised we consider the worst redeemable scenario where
two of the guards are compromised (i.e., $g=2/3$, as $g=1$ is a hopeless scenario). In such case, we can cap $N$ to the
fixed value of 10 using the above equation.

Now, lets look at how to set the value of parameters $K$ and $\mathit{Th}$. For concreteness, we will consider $t=0.2$
(typically this value is assumed in any Tor security model). In the presence of selective DoS, the expected number of
honest and compromised circuits passing the first phase of our detection algorithm are:
$n(HHH)=N\times\frac{(1-g)(1-t)^2}{gt+(1-g)(1-t)^2}$ and $n(CXC)=N\times\frac{gt}{gt+(1-g)(1-t)^2}$ respectively. For
$t=0.2$, $n(CXC)<n(HHH)$ except for $g=1$. So, from an honest circuit's point of view if we were to choose $K=m\cdot
n(CXC)$ (where $m\in\Re, m>0$) probes in the 2nd phase then in the worst case we would have $n(CXC)$ out of the $K$
\textit{candidate} circuits of form $\mathit{CXC}$. In order to have at least 50\% of the probes to be successful, the
following condition must be met: $2n(CXC) < m\cdot n(CXC) < N\Rightarrow 2< m < 1+\frac{(1-g)(1-t)^2}{gt}$. Now, the
best possible outcome that a compromised circuit can achieve is: $n(CXC)-1$ successful probes out of $K$ probes. On the
other hand, the worst possible outcome for an honest circuit is $(m-1)\times n(CXC)$ successful probes out of $K$
probes. With $m>2$, it is clear that $(m-1)\times n(CXC)>n(CXC)$. The value of $\mathit{Th}$ should, therefore, lie in
the range of $(m-1)\times n(CXC)\leq Th<K$. Thus, given the value of $t$, $g$ and $N$ we can determine the range of $K$
and $\mathit{Th}$.

\section{Tuning $K$ and $\mathit{Th}$ using crossover points between $FN$ and $FP$}
\begin{figure}[!h]
\centering
\begin{tabular}{cc}
\epsfig{file=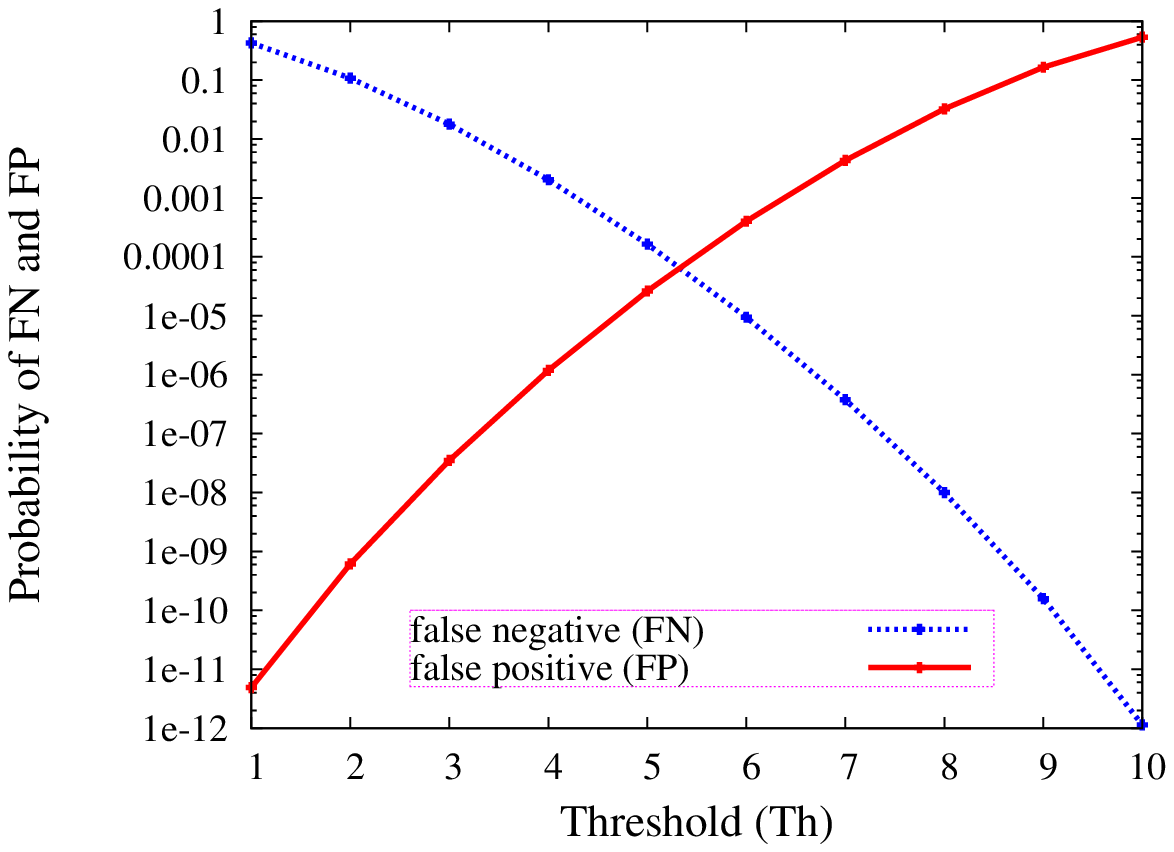,width=0.4\linewidth,clip=}& \epsfig{file=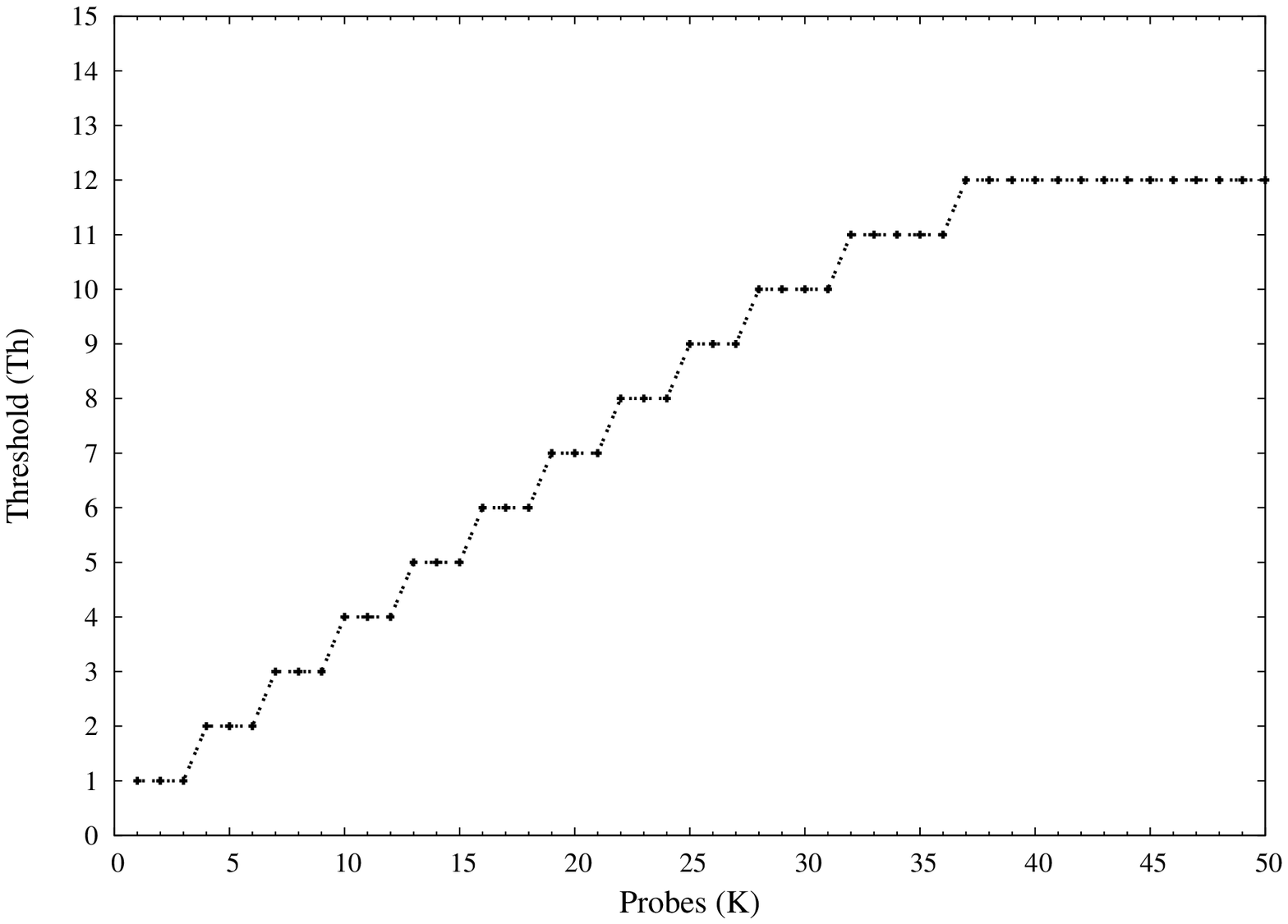,width=0.4\linewidth,clip=}\\
(a)&(b)
\end{tabular}
\caption{(a) Computing the value of $\mathit{Th}$ at which $FN$ and $FP$ crossover. We can see that as $\mathit{Th}$
increases $FP$ goes up while $FN$ goes down. (b) Using this crossover point we can determine the value of $\mathit{Th}$
for a given $K$.}\label{fp-fn-crossover}
\end{figure}

{\label{fn-fp-crossover}}
Parameter $\mathit{Th}$ impacts both $FN$ and $FP$ rating for a given $K$ value. If we increase $\mathit{Th}$ (for a
given $K$) it lowers $FN$ while it increases $FP$. Figure \ref{fp-fn-crossover}(a) shows the probability of $FN$ and
$FP$ against threshold $\mathit{Th}$ for the parametric setting $(t,g,f,d)=(0.2,1/3,0.23,1)$ with $K=10$. The
\textit{y}-axis is given in log-scale. As we can see from the figure the $FN$ and $FP$ crossover and this crossover
point can be used to determine the value of $\mathit{Th}$ to use. In this case (for $K=10$), we see that $\mathit{Th}$
can be set to either 5 or 6 (which conforms with the range of $1\leq Th<K$, computed in Appendix \ref{pr-tunning}). We
can then use this crossover points to compute the pair $(K,\mathit{Th})$. Figure \ref{fp-fn-crossover}(b) also shows
the corresponding values of $K$ and $\mathit{Th}$ at the crossover points. This can be used to tune the value of $K$
and $\mathit{Th}$. \vspace{-2pt}

\end{document}